\documentclass[11pt]{article}

\usepackage{bm}
\usepackage{amsmath}
\usepackage{mathtools}
\usepackage{graphicx}
\usepackage{cite}
\usepackage[margin=1.25in]{geometry}
\usepackage{booktabs}
\usepackage{caption}
\usepackage{subcaption}
\usepackage{nicefrac}
\usepackage{placeins}
\usepackage{afterpage}
\usepackage{float}

\usepackage[bottom]{footmisc}
\usepackage{color}
\definecolor{olivedrab}{rgb}{0.42,0.56,0.14}
\definecolor{oxfordblue}{rgb}{0.0, 0.13, 0.28}

\newcommand{\tensr}[1]{\bm{\mathsf{#1}}}
\newcommand{\um}{\scalebox{0.75}[1.0]{\( - \)}}

\newcommand{\sss}{\scriptscriptstyle}
\newcommand{\sbs}[1]{_{\sss#1}}
\newcommand{\sps}[1]{^{\sss#1}}

\newcommand{\ux}{u\sbs{x}}
\newcommand{\uy}{u\sbs{y}}

\newcommand{\cscs}{c\sbs{s}\sps{2}}

\newcommand{\phiA}{\phi\sbs{a}}
\newcommand{\phiB}{\phi\sbs{b}}
\newcommand{\phiM}{\phi\sbs{m}}
\newcommand{\phiO}{\phi\sbs{o}}

\title{\vspace{-2.0cm}A Robust Lattice Boltzmann Method for Interface-Bound Transport of a Passive Scalar: Application to Surfactant-Laden Multiphase Flows\footnote{Investigations on this research were presented at 32nd International Conference on Discrete Simulation of Fluid Dynamics (DSFD), Albuquerque, New Mexico, July 2023 and at American Physical Society (APS) 76th Annual Division of Fluid Dynamics (DFD) Meeting, Washington D.C., November 2023 (https://meetings.aps.org/Meeting/DFD23/Session/T16.6).}}
\author{{William Schupbach, Kannan Premnath}\\Department of Mechanical Engineering\\ College of Engineering, Design and Computing\\University of Colorado Denver\\1200 Larimer Street, Denver, CO 80204, U.S.A.}

\begin{document}

\maketitle

%%%%%%%%%%%%%%%%%%%%%%%%%%%%%%%%%%%%%%%%%%%%%%%%%%%%% ----------------- ABSTRACT
\begin{abstract}
\noindent
The transport of a passive scalar restricted on interfaces, which is advected by the fluid motions have numerous applications in multiphase transport phenomena. A prototypical example is the advection-diffusion of the concentration field of an insoluble surfactant along interfaces. A sharp-interface model of the surfactant transport on the interface (Stone, Phys. Fluids A, 111, 1990) has been modified and further extended to a diffuse-interface formulation based on a delta-function regularization by Teigen et al. (in \emph{Comm. Math. Sci.}, 4:1009, 2009). However, the latter approach involves singular terms which can compromise its numerical implementation. Recently, Jain and Mani (in \emph{Annual Research Briefs}, Center for Turbulence Research, Stanford University, 2022) circumvented this issue by applying a variable transformation, which effectively leads to a generalized interface-bound scalar transport equation with an additional interfacial confining flux term that prevents lateral diffusion into the bulk phase regions. The resulting formulation has similarities with the conservative Allen-Cahn equation (CACE) used for tracking of interfaces. In this paper, we will discuss a novel robust central moment lattice Boltzmann (LB) method to simulate the interface-bound advection-diffusion transport equation of a scalar field proposed in Teigen et al. by applying Jain and Mani's transformation. It is coupled with another LB scheme for the CACE to compute the evolving interfaces, and the resulting algorithm is validated against some benchmark problems available in the literature. As further extension, we have coupled it with our central moment LB flow solver for the two-fluid motions, which is modulated by the Marangoni stresses resulting from the variation of the surface tension with the local surfactant concentration modeled via the Langmuir isotherm. This is then validated by simulating insoluble surfactant-laden drop deformation and break-up in a shear flow at various capillary numbers.

\end{abstract}

%%%%%%%%%%%%%%%%%%%%%%%%%%%%%%%%%%%%%%%%%%%%%%%%%%%%% ----------------- INTRODUCTION
\newpage
\section{Introduction}
Fluid flows in which a combination of materials existing in multiple thermodynamic phases are classified as multiphase flow, such as the flow of water entrained with water vapour bubbles that often occur in industrial applications such as power generation, water treatment systems, the production of oil, as well as refrigeration and air conditioning systems and in emerging microfluidics. On the other hand, these phase components can also consist of chemically distinct matter such as oil drops in water, and are sometimes classified as multi-fluid flows, which occur in emulsification and flotation processes that have numerous industrial applications. In general, multiphase flows contain a continuous phase and a dispersed phase that are separated by an interface in which inter-molecular force balances between the bulk fluids manifest as surface tension forces at continuum scales.

There exist an interesting class of situations where the advection-diffusive transport of a scalar field is confined by the dynamics of the interfaces, with the surface active agents or surfactants being a prototypical example. As such, surfactants are molecules with a hydrophilic head and a hydrophobic tail that tend to preferentially adsorb on fluid interfaces that further disrupt the existing imbalance of inter-molecular forces at the interface and tend to decrease the surface tension locally, which in turn can influence the dynamics of bulk fluids significantly~\cite{manikantan2020surfactant}. There are a wide variety of surface acting agents that exist, both natural and synthetic, but they can be broadly characterized by the head group polarity which leads to a complex range of adsorption and solution properties. The most commonly observed condition for the use of surface active agents, is when the hydrophilic portion is separated into the aqueous phase while the hydrophobic part adsorbs onto a liquid interface. When this occurs, the surfactant effectively acts as a wetting or dispersion agent, as an emulsifier, as a foaming or anti-foaming agent, or as a lubricant such as those found in detergents, pharmaceuticals, and cosmetics, and many other industrial processes. Of interest in this work, is the so called insoluble surfactant, which can be thought of as a surfactant with weak solubility in the bulk phase, while the hydrophobic part of the tail-head edifice still plays an active role in adsorbing to liquid interfaces as they undergo surface diffusion and advection and thereby modifying the surface tension locally.

Due to the complex nature of such surfactant-laden multiphase flows, studying the associated fluid dynamics requires experimental investigations or numerical simulations with appropriate continuum models for their transport. Stone~\cite{stone1990simple} derived a time dependent advection-diffusion equation for interfacial surfactant transport by assuming a `sharp' interface model~\cite{stone1990simple}, where the interface serves as a discontinuity for the bulk fluids with zero thickness. This theoretical framework was then used in simulations using a variety of numerical techniques, such as the boundary integral method~\cite{stone1990effects}, volume-of-fluid method~\cite{james2004surfactant, renardy2002new}, level set method~\cite{xu2012level}, finite element method~\cite{pozrikidis2004finite}, and finite difference method~\cite{khatri2011numerical}.

On the other hand, diffuse-interface models, which are endowed with interfaces of finite thickness across which fluid properties vary continuously and also known as the phase field models, have been used widely in recent years for simulation of multiphase flows (see e.g.,~\cite{anderson1998diffuse,yue2004diffuse,ding2007diffuse}) as they overcome numerical challenges associated with interface tracking with methods that are based on sharp-interface models. Many different formulations to incorporate the effect of surfactants in such phase field models have been developed in recent years (see e.g.,~\cite{van2006diffuse,liu2010phase, engblom2013diffuse,van2014mesoscale,van2016analysis,shi2019improved,soligo2019coalescence}). In the context of modeling insoluble surfactants with diffuse interfaces, Teigen et al.~\cite{teigen2009diffuse} modified the formulation of Stone~\cite{stone1990simple} by introducing a surface delta function, which was solved by various different numerical schemes, including the lattice Boltzmann method (LBM) recently~\cite{liu2018hybrid,hu2021diffuse}. However, the Teigen et al.~\cite{teigen2009diffuse}'s approach was found to have singularity issues as they involve computation of the inverse of the surface delta functions outside the interface region. Recognizing this, Jain and Mani~\cite{jainmodeling} recently circumvented such an issue via a variable transformation based on their computational model for transport of immiscible scalars in two-phase flows~\cite{jain2023computational}. This leads to a regularized formulation to represent the advection-diffusion of interface-confined scalars and interestingly has a mathematical structure similar to the conservative Allen-Cahn equation (CACE) used for capturing interfaces~\cite{chiu2011conservative}. It forms the basis for further extensions and numerical implementations using an advanced lattice Boltzmann (LB) approach in this paper.

The lattice Boltzmann method (LBM)~\cite{benzi1992lattice,lallemand2021lattice} is a computational approach based on kinetic theory and involves collide-and-stream steps of the particle distribution functions on a discrete lattice. Due to algorithmic simplicity and natural parallelization capabilities as well as its ability to serve as a rich platform to incorporate models based on kinetic theory at mesoscopic scales, it has been used for simulations of a wide range of fluid mechanics applications, including multiphase flows. The choice of the collision model in LBM plays a major role in its fidelity and numerical stability. In particular, the central moment-based formulations such as the cascaded model utilizing the Maxwell distribution~\cite{geier2006cascaded} and the Fokker-Planck model~\cite{schupbach2024fokker} have been shown to result in significantly improved numerical characteristics with robust implementations. Moreover, the central moment-based LB approach has recently been used for interface capturing and simulation of multiphase flows (see e.g.,~\cite{hajabdollahi2021central}).
Thus, in this paper, we develop a new central moment LB method for the transport of an interface-bound scalar field representing the insoluble surfactant concentration. Moreover, as a further extension of~\cite{jainmodeling}, we couple this approach with two other LB solvers based on central moments for interface capturing and the computation of two-fluid motions with variable surface tension effects; the latter involve the normal capillary and the tangential Marangoni forces, and the dependence of the local surface tension on the surfactant concentration is parameterized via the Langmuir isotherm. The resulting numerical schemes are tested and validated for a series of benchmark problems to demonstrate their accuracy in simulations of surfactant-laden multiphase flows.

The outline of this paper is as follows. In the next section (Sec.~\ref{sec:goveqns}), we discuss the derivation of the regularized interface-confined transport of a scalar field, as well as the governing equations for CACE-based interface capturing and two-fluid motions with a variable surface tension modeling. Next, Sec.~\ref{sec:centralmomentLBM} presents the development of a novel central moment LBM for solving the interface-confined scalar transport equation, while the other two LB solvers for interface capturing and two-fluid motions are discussed in the appendix sections. Then, the results for a series of validation benchmark cases are discussed in Sec.~\ref{sec:resultsandconclusions}. Finally, a summary of this paper and the main conclusions drawn are presented in Sec.~\ref{sec:summaryandconclusions}.

%%%%%%%%%%%%%%%%%%%%%%%%%%%%%%%%%%%%%%%%%%%%%%%%%%%%% ----------------- MODELING EQUATIONS
\section{Governing Equations}\label{sec:goveqns}
We now discuss the essential elements of the development of the interface-restricted transport of the scalar field, viz., the concentration of an insoluble surfactant, that was proposed by Jain and Mani~\cite{jainmodeling}. Furthermore, we also outline the governing equations of the evolution of diffuse interfaces represented by the conservative Allen-Cahn equation and the multiphase Navier-Stokes equations written in a single-field formulation with variable surface tension forces, with the latter being modulated by the local surfactant concentration field through the Langmuir isotherm discussed at the end of the section. These governing equations and models then form the basis of the development of our novel LB schemes based on central moments in the next section.

\subsection{Interface-bound scalar transport equation for diffuse interfaces}
The ``sharp" interface-based conservation equation of the interfacial surfactant concentration field, which serves as the starting point of this discussion, was first derived by Stone~\cite{stone1990simple} and also elaborated further in~\cite{wong1996surfactant}, is written as
\begin{equation}
\frac{\partial \hat{c}}{\partial t} + \bm{u}\cdot \bm{\nabla}\hat{c} = \bm{\nabla\!\!\sbs{s}}\cdot(D\sbs{s}\bm{\nabla\!\!\sbs{s}} \hat{c}) - \hat{c} \bm{\nabla\!\!\sbs{s}}\cdot\bm{u\sbs{s}} - \hat{c}\bm{u}\cdot\bm{\hat{n}},
\label{surf}
\end{equation}
\noindent
where $\hat{c}$ is the interfacial surfactant concentration,  $\bm{\hat{n}}$ is the unit normal of the interface, $\bm{\nabla\!\!\sbs{s}} = (I-\bm{\hat{n}}\bm{\hat{n}})\bm{\nabla}$ is the surface gradient operator, $D\sbs{s}$ is the interfacial surfactant diffusion coefficient and $\bm{u}$ is a velocity field. This equation represents the advection and diffusion of surfactant concentration $\hat{c}$ exclusively on the interface, which is henceforth denoted by $\gamma$.

\begin{figure}[H]
\centering
\begin{subfigure}{0.25\textwidth}
\includegraphics[trim = 280 100 280 100, clip, width =40mm]{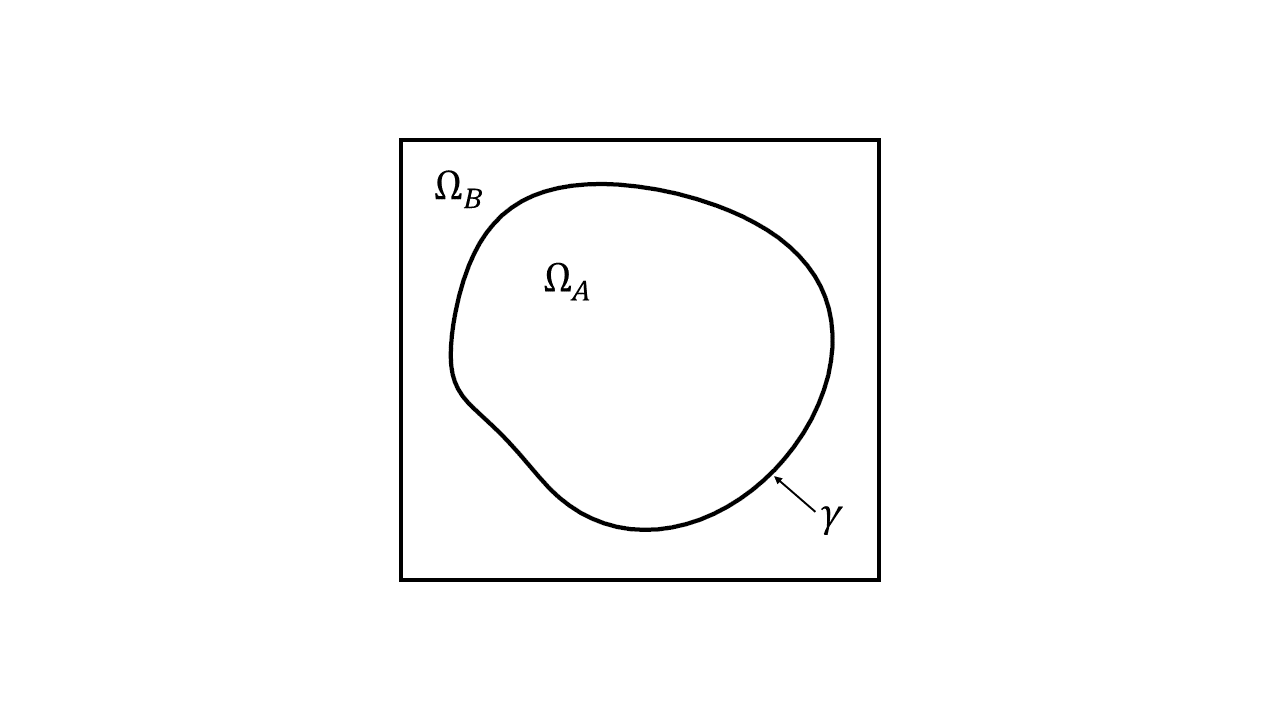}
\end{subfigure}
\caption{Schematic of an interface $\gamma$ separating two fluid domains $\Omega_A$ and $\Omega_B$, which, when combined becomes the total domain $\Omega$, i.e., $\Omega=\Omega_A \cup \Omega_B $.}
\label{potato2}
\end{figure}

On the other hand, for practical implementations based on phase field models, we need the equivalent formulation of Eq.~(\ref{surf}), which was developed by Teigen et al~\cite{teigen2009diffuse} via using the surface delta function $\delta\sbs{s}$ that effectively distributes the scalar field across a narrow region spanning the``diffuse" interfaces. More precisely, this is accomplished by satisfying the conservation relation given by
\begin{equation}
\int\sbs{\gamma}\hat{c} \, d\gamma = \int\sbs{\Omega} \hat{c}\delta\sbs{s} \, d\Omega,
\end{equation}
\noindent
where $\gamma$ represents the interface, and $\Omega$ represents the overall domain (see Fig.~\ref{potato2}).
\noindent
Physically, this means that the total amount of surfactant on the interface is equivalent to the total amount found in the extended domain where such an extension is achieved via the surface delta function. The resulting formulation, which is valid everywhere in the domain $\Omega$ is
\begin{equation}
\frac{\partial}{\partial t} (\hat{c} \delta\sbs{s}) + \bm{\nabla} \cdot (\bm{u} \hat{c} \delta\sbs{s}) = \bm{\nabla}\cdot (D\sbs{s} \delta\sbs{s}\bm{\nabla}\hat{c}).
\label{surfdist}
\end{equation}
\noindent
Then again, Eq.~(\ref{surfdist}) leads to robustness issues since its solution leads to $\hat{c}\delta\sbs{s}$ and to recover $\hat{c}$, one needs to divide $\hat{c}\delta\sbs{s}$ by $\delta\sbs{s}$, which is singular away from the interface.

\subsection{Regularized interface-bound scalar transport equation for diffuse interfaces}
Alternatively, Jain and Mani \cite{jainmodeling}, proposed to regularize the formulation given in Eq.~(\ref{surfdist}) by introducing a transformation of variables $c=\hat{c}\delta\sbs{s}$. Then, Eq.~(\ref{surfdist}), in terms of the new variable $c$, becomes
\noindent
\begin{equation}
\frac{\partial c}{\partial t} + \bm{\nabla}\cdot (\bm{u}c) = \bm{\nabla} \cdot \left[ D\sbs{s} \left( \bm{\nabla}c - \frac{c}{\delta\sbs{s}} \bm{\nabla}\delta\sbs{s}\right)\right].
\label{surf_delta}
\end{equation}
In what follows, we consider that the interface $\gamma$ is tracked by the conservative Allen-Cahn equation (CACE) for the phase field variable $\phi$ with bounds $\phi\sbs{b} \leq \phi \leq \phi\sbs{a}$, which is given by~\cite{chiu2011conservative}
\begin{equation}
\frac{\partial \phi}{\partial t} + \bm{\nabla}\cdot\left(\bm{u}\phi\right) = \bm{\nabla}\cdot \left[ M\sbs{\phi}\left(\bm{\nabla}\phi - \theta \bm{\hat{n}} \right)\right],
\label{CACE}
\end{equation}
where $\bm{\hat{n}} = \bm{\nabla}\phi / |\bm{\nabla}\phi|$ is the unit normal, $\phi=\phi\sbs{a}$ in $\Omega_A$ and $\phi=\phi\sbs{b}$ in $\Omega_B$, and the coefficient $\theta$ is given by
\begin{equation}
\theta = \left( \frac{4}{W}\right)\frac{(\phiA-\phi)(\phi-\phiB)}{(\phiA-\phiB)}. \nonumber
\label{theta}
\end{equation}
Here, $W$ is the width of the interface. Then, an equilibrium profile across the  interface can be derived from Eq.~(\ref{CACE}) by letting the interfacial diffusion term balance with the sharpening term, i.e., $M\sbs{\phi}\left(\bm{\nabla}\phi - \theta \bm{\hat{n}}\right)=0 $ or $\bm{\nabla}\phi = \theta \bm{\hat{n}}$, and the resulting hyperbolic tangent profile reads as
\begin{equation}
\phi\sps{eq} = \phi\sbs{m} + \phi\sbs{o} \tanh{\left(\frac{2\zeta}{W}\right)},
\label{CACEeq}
\end{equation}
where $\phiM = (\phiA+\phiB)/2$ and $\phiO = (\phiA-\phiB)/2$, $\zeta$ is the interfacial normal coordinate with $\zeta = 0$ where $\phi=\phi_m$. Then, following~\cite{teigen2009diffuse}, we obtain the surface delta function $\delta\sbs{s}$ via $\delta\sbs{s} = |\bm{\nabla} \phi|$, and in view of the above, and noting $|\bm{\hat{n}}|=1$, it follows that
\begin{equation}
\delta\sbs{s} = |\bm{\nabla}\phi| = \theta.
\label{delta}
\end{equation}
Then, we can crucially observe that
\begin{equation}
\frac{1}{\delta\sbs{s}}\bm{\nabla}\delta\sbs{s}=\frac{1}{\delta\sbs{s}}\frac{d\theta}{d\phi}\bm{\nabla}\phi=\frac{d\theta}{d\phi}\bm{n}
\end{equation}
Finally, using this last equation in the second term on the right hand side of Eq.~(\ref{surf_delta}) and simplifying we arrive at the regularized form of the interface-confined scalar transport equation, which reads as
\begin{equation}
\frac{\partial c}{\partial t} + \bm{\nabla}\cdot \left( \bm{u} c\right) =\bm{\nabla}\cdot \left[ D\sbs{s} \left( \bm{\nabla} c  -\theta\sbs{c}\bm{\hat{n}} \right)  \right],
\label{ISS}
\end{equation}
where the coefficient $\theta\sbs{c}$ is given by
\begin{equation}
\theta\sbs{c} = \frac{8 c}{W}\frac{ ( \phiM - \phi ) }{(\phiA-\phiB)}.
\end{equation}
Here, the term involving $D\sbs{s}\theta\sbs{c}\bm{n}$ in Eq.~(\ref{ISS}) can be interpreted as the ``confining" flux term, which prevents the diffusion of the scalar field (i.e, surfactant) on both sides of the interface and restricts it exclusively to the interface region. It is important to point out that Eq.~(\ref{ISS}) does not require dividing by $\delta\sbs{s}$ and avoids the singularity issue associated with Eq.(\ref{surfdist}), and is therefor more robust. Moreover, unlike Eq.~(\ref{surf}), there is no need to deal with the surface gradient operator $\bm{\nabla\sbs{\!\!s}}$, and hence Eq.~(\ref{ISS}) is simpler to use in numerical implementation. Equation~(\ref{ISS}) represents a slight generalization of the model derived in~\cite{jainmodeling}, which uses $0 \leq \phi \leq 1$ whereas here we have $\phiB \leq \phi \leq \phiA$.

We note here that the equilibrium concentration profile across the interface $c=c(\zeta)$ can be readily obtained from Eq.~(\ref{ISS}) using the same approach as that used to derive Eq.~(\ref{CACEeq}) and is given here as~\cite{jainmodeling}
\begin{equation}
c = \frac{c_0}{\cosh^2\left(\frac{2\zeta}{W}\right)}.
\label{ISSeq}
\end{equation}
Here, $c_0$ is the reference concentration field at the interface where $\zeta = 0$ and very rapidly decays to zero away from the diffuse interface region with $c=0$ as $\zeta \rightarrow \pm\infty$. In practice, we use Eq.~(\ref{ISSeq}) for the initialization of the surfactant concentration field $\hat{c}$ in simulations.

\subsection{Navier-Stokes equations for multiphase flows with variable surface tension forces}
In order to simulate more general situations where both the interface and the concentration field is advected by the motion of two-fluids in multiphase flows, thereby extending the recent work~\cite{jainmodeling}, we need to be able to compute the hydrodynamic velocity fields. In this regard, within the phase field framework, we utilize the single-field formulation of the Navier-Stokes equations with a distributed form of surface tension forces. Thus, the corresponding mass and momentum equations are given, respectively, as
\begin{equation}
\frac{\partial \rho}{\partial t} + \bm{\nabla}\cdot (\rho \bm{u}) = 0,
\end{equation}
and
\begin{equation}
\frac{\partial (\rho \bm{u})}{\partial t} + \bm{\nabla}\cdot (\rho \bm{u}\bm{u}) = -\bm{\nabla}P + \bm{\nabla}\left[\mu\left(\bm{\nabla}\bm{u} + \bm{\nabla}\bm{u}\sps{\top}\right)\right] + \bm{F\sbs{t}},
\end{equation}
where the total force $\bm{F\sbs{t}}$ consists of the combination of the surface tension force $\bm{F\sbs{s}}$ and any external body force $\bm{F\sbs{ext}}$ given as
\begin{equation}
\bm{F\sbs{t}} = \bm{F\sbs{s}}  + \bm{F\sbs{ext}},
\end{equation}
and $\bm{u}$ and $P$ are the fluid velocity and pressure, respectively. Here, the density $\rho$ and viscosity $\mu$ are taken to vary smoothly across the diffuse interfaces.

\subsubsection{Modeling of variable surface tension force}
\begin{figure}[H]
\centering
\begin{subfigure}{0.48\textwidth}
\includegraphics[trim = 0 0 0 0, clip, width =70mm]{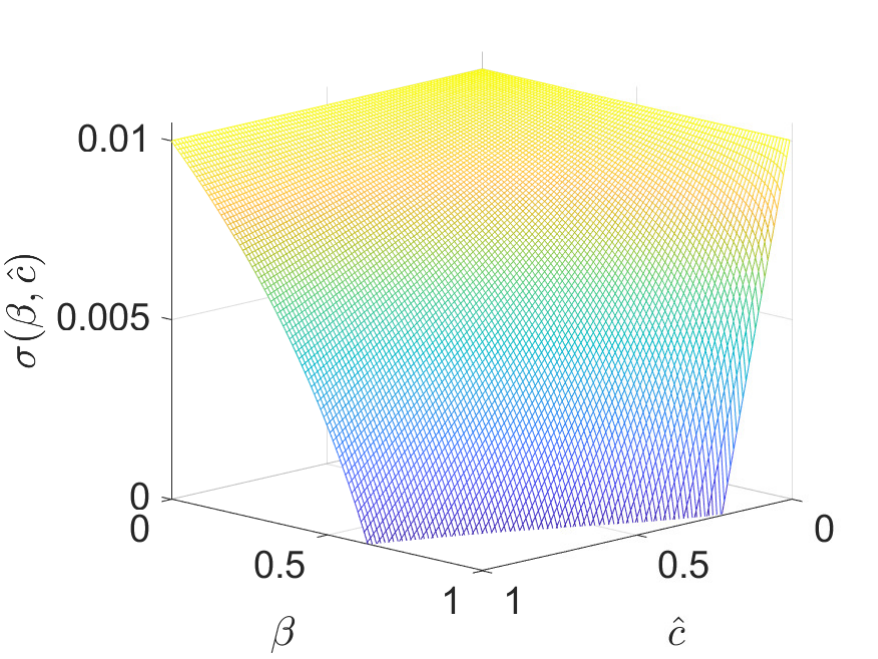}
\label{Langmuir_Isotherm}
\end{subfigure}
\caption{Illustration of the Langmuir Isotherm surface indicating that the local surface tension decreases for any given values of Gibbs elasticity parameter and surfactant concentration when compared to the clean interface case.}
\label{LangmuirIso}
\end{figure}

In the presence of surfactants on the interface, the surface tension force, shown below in Eq.~(\ref{Fs}), contains both the capillary force, which is normal to interface, and the Marangoni force, which is tangential to the interface, and they depend on the local interfacial concentration $c$. Thus, we can resolve the distributed form of the local surface tension force as
\begin{equation}
\bm{F\sbs{s}}= \underbrace{-\sigma(c) |\bm{\nabla}\phi |\sps{2} \left(\bm{\nabla}\cdot\bm{n}\right)\bm{n}}_{\mbox{capillary force}} +  \underbrace{|\bm{\nabla}\phi |\sps{2}\bm{\nabla\sbs{s}}\sigma(c)}_{\mbox{Marangoni force}}
\label{Fs}
\end{equation}
The local surface tension $\sigma(c)$ dependence on the surfactant concentration is modeled by means of a surface equation of state given by the standard Langmuir isotherm equation of state, which reads as
\begin{equation}
\sigma(c) = \sigma\sbs{o}\left[ 1 + \beta \log \left(1-\frac{c}{c\sbs{max}} \right)\right],
\label{EOS}
\end{equation}
where $\sigma\sbs{o}$ is the surface tension for the clean interface, i.e., in the absence of surfactants, $\beta$ is the Gibbs elasticity parameter, which represents the sensitivity of surface tension on the surfactant concentration, and $c\sbs{max}$ is the maximum possible concentration of the latter. Figure~\ref{LangmuirIso} shows an illustration of the Langmuir isootherm surface for ranges of the surfactant concentration and the Gibbs elasticity parameter for $\sigma\sbs{o}=0.01$. It also indicates that there exists certain ranges of parameters that can make the surface tension become negative; in this work, we avoid choosing the range of parameters that result in negative values of the local surface tension.

%%%%%%%%%%%%%%%%%%%%%%%%%%%%%%%%%%%%%%%%%%%%%%%%%%%%% ----------------- LBM APPROACH
\section{Central moment LBM for Transport of Interface-Bound Scalar Field} \label{sec:centralmomentLBM}
The governing equations given in the previous section represent a coupled system for 1) the transport of the interface-confined scalar field, 2) the tracking of interfaces, and 3) computation of multiphase flows with variable surface tension forces. In this paper, we develop novel central moments-based LB schemes using the D2Q9 lattice for computing each of these three components. They each involve computing a separate LB equation whose collision is modeled by relaxation of nine central moments of the respective distribution functions to their carefully constructed equilibria and followed by the streaming of the resulting post-collision distribution functions; and from the latter, the macroscopic variable of interest, viz., the concentration field $c$, order parameter $\phi$, or the hydrodynamic velocity field $\bm{u}$, are updated. Since the computation of $c$ depends on $\phi$ and $\bm{u}$, and the latter, in turn, is modulated by $c$ via the Langmuir isotherm for the local surface tension, their solutions are fully coupled. Among these, the focus of this paper is on the development of a new central moment LB scheme for the interface-confined scalar field for simulating the transport of the insoluble surfactants, which is discussed in this section, while the other two LB schemes are extensions of our recent work~\cite{hajabdollahi2021central,elbousefi2023thermocapillary,elbousefi2024investigation} and hence they are presented in various appendices for completeness.

%The lattice Boltzmann equation, shown in Eq.~(\ref{LBE}), is used to obtain a solution to Eq.~(\ref{ISS}), by evolving the distribution function $h\sbs{\alpha}$, where $\alpha = 0,1,2\dots,8$ representing the discrete particle velocity directions associated with the $d2q9$ lattice.

\begin{figure}[H]
\centering
\includegraphics[trim = 0 0 0 0, clip, width =40mm]{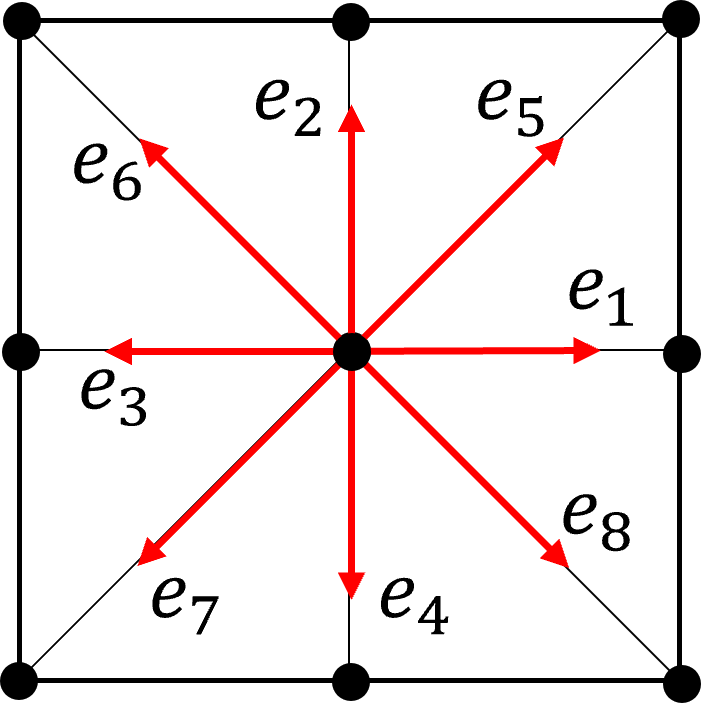}
\caption{A schematic of the two-dimensional, nine particle velocity (D2Q9) lattice used to develop the LB algorithms in this manuscript.}
\label{d2q9lattice}
\end{figure}

The discrete distribution functions for the evolution of the interface-restricted scalar field $c$ is given by $h\sbs{\alpha}$, where $\alpha = 0,1,2\dots,8$ are the nine discrete particle velocity directions $\bm{e}_\alpha$ associated with the D2Q9 lattice (see Fig.~\ref{d2q9lattice}). In general, the simplest model for collision represents the relaxation of $h\sbs{\alpha}$ towards their respective equilibrium distributions $h\sbs{\alpha}\sps{eq}$. Instead, in this work, we introduce an advanced formulation that relaxes the central moments of the distribution functions to their appropriate equilibria so that the transport of $c$ as prescribed by Eq.~(\ref{ISS}) is recovered in simulations.
\noindent

The Cartesian components of the velocity directions associated with this lattice can be expressed as follows:
\begin{subequations}
\begin{equation}
\qquad \left| \bm{e}_x \right> = ( 0, 1, 0, -1, 0, 1, -1, -1, 0)\sps{\top},
\label{ex}
\end{equation}
\begin{equation}
\qquad \left| \bm{e}_y \right> = ( 0, 0, 1, 0,-1, 1, 1, -1, -1)\sps{\top}.
\label{ey}
\end{equation}
\end{subequations}
Here and henceforth, the superscript $\top$ refers to the transpose operator. We use the standard Dirac's bra-ket notation, that is $\left<\cdot\right|$ and $\left|\cdot\right>$, represent the row and column vectors, respectively. Then, the nine distribution functions can be collected as the vector $\left|\mathbf{h}\right>=(h\sbs{0},h\sbs{1},\ldots h\sbs{8})\sps{\top}$. For constructing the LB scheme, we also need the following 9-component vector containing each element equal to unity given by
\begin{eqnarray} %\label{eqn34}
\qquad \left|\mathbf{1}\right> = (1,1,1,1,1,1,1,1,1)\sps{\top}. \nonumber
\end{eqnarray}
so that its inner product with the vector of distribution functions $\mathbf{h}$, i.e., $\left<\mathbf{h}|\mathbf{1}\right>$ will yield the local interfacial concentration field: $c=\left<\mathbf{h}|\mathbf{1}\right>$. Then, the nine independent non-orthogonal basis vectors used to construct a central moment LB formulation in what follows are given by
\begin{gather}
\qquad \left| \mathbf{p}_0 \right> = \left| \mathbf{1} \right>, \quad
\left| \mathbf{p}_1 \right> = \left| \bm{e}_x \right>, \quad
\left| \mathbf{p}_2 \right> = \left| \bm{e}_y \right>, \nonumber \\[2mm]
\qquad \left| \mathbf{p}_3 \right> = \left| \bm{e}_x^2\right>, \quad
\left| \mathbf{p}_4 \right> = \left| \bm{e}_y^2\right>, \quad
\left| \mathbf{p}_5 \right> = \left| \bm{e}_x \bm{e}_y \right>,\nonumber \\[2mm]
\qquad \left| \mathbf{p}_6 \right> = \left| \bm{e}_x^2 \bm{e}_y \right>,\quad
\left| \mathbf{p}_7 \right> = \left| \bm{e}_x \bm{e}_y^2 \right>,\quad
\left| \mathbf{p}_8 \right> = \left| \bm{e}_x^2 \bm{e}_y^2 \right>.
\label{Pi}
\end{gather}
These can be utilized for mapping the distribution functions to the respective nine raw moments via taking the inner products of $\left<\mathbf{h}\right|$ with $\left|\mathbf{p}_j\right>$, where $j=0,1,\ldots,8$. Such operations can be compactly represented by grouping together all the basis vectors as the following transformation matrix $\tensr{P}$:
\noindent
\begin{equation}
\qquad \tensr{p} = \left[
\left|\mathbf{p}_0\right>,
\left|\mathbf{p}_1\right>,
\left|\mathbf{p}_2\right>,
\left|\mathbf{p}_3\right>,
\left|\mathbf{p}_4\right>,
\left|\mathbf{p}_5\right>,
\left|\mathbf{p}_6\right>,
\left|\mathbf{p}_7\right>,
\left|\mathbf{p}_8\right>
\; \right]\sps{\top},
\label{P}
\end{equation}
For convenience, the elements of this transformation matrix and its inverse are given in Appendix~\ref{A}. Then, the discrete raw moments of the distribution functions $\chi'\sbs{mn}$ and the equilibrium distribution functions $\chi^{\;\sss{eq'}}\sbs{mn}$ of order ($m+n$) can be defined in their component forms as follows:
\begin{equation}
\qquad \left( \begin{array}{c}\chi'\sbs{mn}\\[2mm]   \chi^{\;\sss{eq'}}\sbs{mn} \end{array} \right)  = \sum\sbs{\alpha = 0}\sps{8} \left( \begin{array}{c}h\sbs{\alpha} \\[2mm]   h\sbs{\alpha}\sps{eq} \end{array} \right)  e\sbs{\alpha x}\sps{m}   e\sbs{\alpha y}\sps{n}.
\label{rm}
\end{equation}
Since our goal is to actually construct a central moments-based LB scheme, next, we define the discrete central moments of $h\sbs{\alpha}$ and its equilibria of order ($m+n$), which read as
\begin{equation}
\qquad \left( \begin{array}{c}\chi\sbs{mn} \\[2mm]   \chi\sbs{mn}\sps{eq} \end{array} \right)  = \sum\sbs{\alpha = 0}\sps{8} \left( \begin{array}{c}h\sbs{\alpha} \\[2mm]  h\sbs{\alpha}\sps{eq} \end{array} \right) (e\sbs{\alpha x}-u\sbs{x})\sps{m}  ( e\sbs{\alpha y}-u\sbs{y})\sps{n}.
\label{cmscalar}
\end{equation}
Then, for convenience, we represent all of the independent discrete raw moments and central moments of various orders, associated with the D2Q9 lattice using the vectors $\bm{\chi'}$ and $\bm{\chi}$, respectively, as
\begin{subequations}
\begin{eqnarray}
\qquad \bm{\chi'} \! \! \! &=& \! \! \! ( \chi'\sbs{00}, \chi'\sbs{10},\chi'\sbs{01}, \chi'\sbs{20}, \chi'\sbs{02}, \chi'\sbs{11},\chi'\sbs{21}, \chi'\sbs{12},\chi'\sbs{22} ),\label{eqn:4a} \\[3mm]
\qquad \bm{\chi} \! \! \! &=& \! \! \! ( \chi\sbs{00},\chi\sbs{10}, \chi\sbs{01}, \chi\sbs{20}, \chi\sbs{02}, \chi\sbs{11}, \chi\sbs{21}, \chi\sbs{12}, \chi\sbs{22} ).
\end{eqnarray}
\end{subequations}
Based on the above considerations, we now note that the distribution functions can be mapped into raw moments by $\bm{\chi'} = \tensr{P}\mathbf{h}$, and can then be mapped further into central moments using $\bm{\chi} = \tensr{F}\bm{\chi'}$, where the frame transformation matrix $\tensr{F}$ is defined by taking binomial expansions of Eq.~(\ref{cmscalar}) (see e.g.,~\cite{yahia2021central}). For convenience, the latter as well as its inverse are presented in Appendix~\ref{B}. Conversely, one can perform the reverse mappings via applying, in turn, the inverse matrices of $\tensr{F}$ and $\tensr{P}$ on central moments $\bm{\chi}$ to obtain the distribution functions $\mathbf{h}$.

\begin{figure}[H]
\centering
\begin{subfigure}{0.48\textwidth}
\includegraphics[trim = 0 0 0 0, clip, width =100mm]{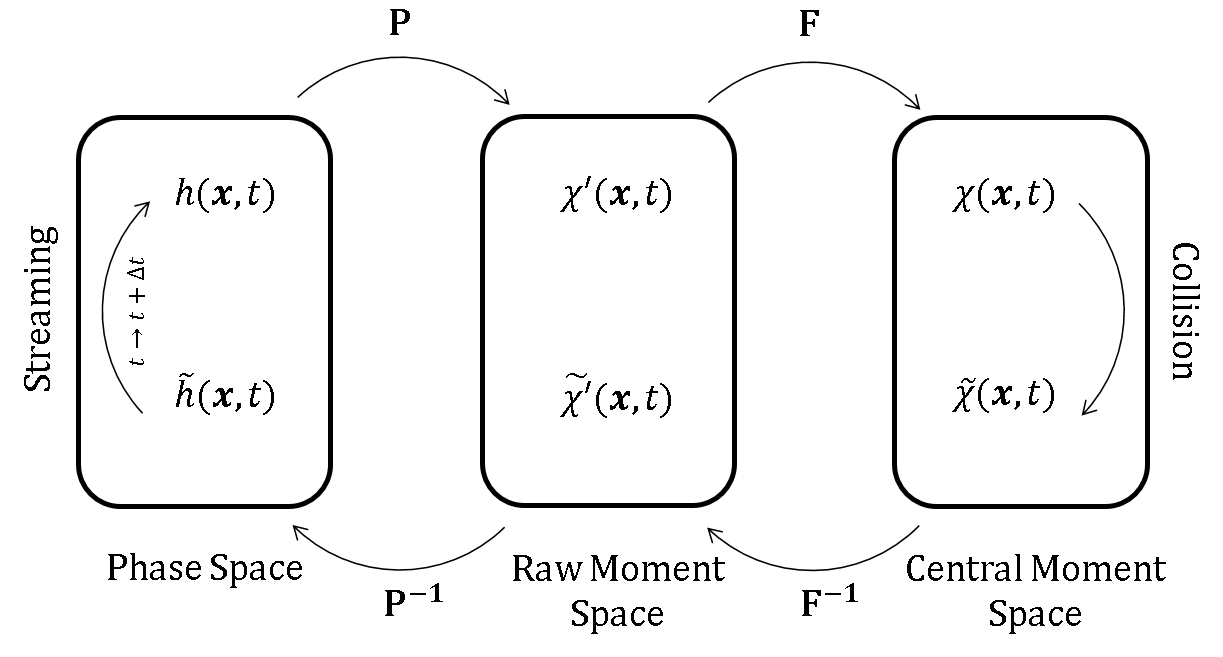}
\end{subfigure}
\caption{Graphical representation of the collision step performed using central moments via using the associated mappings between the distribution functions and central moments, which is followed by the streaming step in terms of the distribution functions.}
\label{fig:LBalgorithm}
\end{figure}

Next, to construct the collision step, we define the discrete central moment equilibria by means of a matching principle whereby we match the independent discrete central moments supported by D2Q9 lattice with the corresponding continuous central moments of the Maxwell distribution, after replacing the density $\rho$ with the insoluble surfactant concentration $c$. Moreover, the flux term in Eq.~(\ref{ISS}) involving $D\sbs{s}\theta\sbs{c}\bm{n}$ is accounted for using the method of extended moment equilibria, where the first order equilibrium central moments are augmented to include these terms. A Chapman-Enskog analysis, not presented here for brevity, confirms that Eq.~(\ref{ISS}) is recovered. The resulting central moment equilibria for the D2Q9 lattice then read as
\begin{gather}
\qquad \chi\sbs{00}\sps{eq} = c, \qquad
\chi\sbs{10}\sps{eq} = D\sbs{s} \theta\sbs{c}  n\sbs{x},\qquad
\chi\sbs{01}\sps{eq} = D\sbs{s} \theta\sbs{c}  n\sbs{y},\nonumber \\[2mm]
\qquad \chi\sbs{20}\sps{eq} = c\sbs{s}\sps{2}\cdot c,\qquad
\chi\sbs{02}\sps{eq} = c\sbs{s}\sps{2}\cdot c,\qquad
\chi\sbs{11}\sps{eq} = 0,\nonumber  \\[2mm]
\qquad \chi\sbs{21}\sps{eq} = 0,\qquad
\chi\sbs{12}\sps{eq} = 0,\qquad
\chi\sbs{22}\sps{eq} = c\sbs{s}\sps{4}\cdot c,
\end{gather}
where $c_{s}^2=1/3$, with $c_s$ being the lattice speed of sound. From the above considerations, we can then summarize the central moment LB scheme to solve the interface-bound surfactant concentration field as follows:
\begin{equation}
h\sbs{\alpha}(\bm{x}+\bm{e}\sbs{\alpha}\Delta t, t+\Delta t) - h\sbs{\alpha}(\bm{x}, t) =  \tensr{P}\sps{\um 1}\tensr{F}\sps{\um 1} \left\{\bm{\Lambda\sbs{\chi}} \left[ \bm{\chi}\sps{eq} - \bm{\chi}\right] \right\},
\label{LBE}
\end{equation}
where $\bm{\Lambda\sbs{\chi}} = \mbox{diag}(1, \omega\sbs{\chi}, \omega\sbs{\chi},1 ,1,1,1,1,1)$ is the diagonal relaxation rate matrix in which the relaxation rate $\omega\sbs{\chi}$ for the first order moments is related to the diffusion coefficient $D\sbs{s}$ by
\begin{equation}\label{eq:diffcoeffscalarfield}
D\sbs{s}=c\sbs{s}\sps{2}\left(1/\omega\sbs{\chi} - 1/2\right)\Delta t.
\end{equation}
What follows next are the steps involved in the algorithm to solve this central moment LBE for one time step $\Delta t$ which is graphically depicted in Fig.~\ref{fig:LBalgorithm}.

\begin{itemize}
\item Compute pre-collision raw moments from distribution functions using
\begin{equation}
\bm{\chi'} = \tensr{P}\bm{h} \nonumber
\end{equation}

\item Compute pre-collision central moments from raw moments via
\begin{equation}
\bm{\chi} = \tensr{F}\bm{\chi'} \nonumber
\end{equation}

\item Perform collision by relaxing the central moments towards equilibrium central moments
\begin{equation}
\bm{\tilde{\chi}} = \bm{\chi} + \bm{\Lambda\sbs{\chi}} \left( \bm{\chi\sps{eq}} - \bm{\chi}\right) ,\nonumber
\end{equation}

\item Compute post-collision raw moments from central moments via
\begin{equation}
\bm{\tilde{\chi}'} = \tensr{F}\sps{-1}\bm{\tilde{\chi}} \nonumber
\end{equation}

\item Compute post-collision distribution functions from raw moments through
\begin{equation}
\bm{\tilde{h}} = \tensr{P}\sps{-1}\bm{\tilde{\chi}'} \nonumber
\end{equation}

\item Perform streaming via
\begin{equation}
h\sbs{\alpha}(\bm{x},t+\Delta t) = \tilde{h}\sbs{\alpha}(\bm{x}-\bm{e}\sbs{\alpha}\Delta t, t)  \nonumber
\end{equation}
where $\alpha = 0,1,2\dots8$.

\item Update the local insoluble surfactant concentration by computing the zeroth moment of the distribution functions $h\sbs{\alpha}$ as
\begin{equation}
c= \sum\sbs{\alpha=0}\sps{8}h\sbs{\alpha}.
\end{equation}
\end{itemize}
In addition, the central moment LBM for interface tracking based on conservative Allen-Cahn equation to compute $\phi$ and another central moment LBM for two-fluid hydrodynamics with variable surface tension effects to obtain $\bm{u}$ are presented in Appendix~\ref{app:CACE} and Appendix~\ref{app:NSE}, respectively.

%%%%%%%%%%%%%%%%%%%%%%%%%%%%%%%%%%%%%%%%%%%%%%%%%%%%% ----------------- RESULTS
\section{Results and Discussion}\label{sec:resultsandconclusions}
We will present a series of benchmark test cases that demonstrate the efficacy of our new central moment LB formulation used to obtain the evolution of the interface-bound surfactant concentration field as modeled by Eq.~(\ref{ISS}). The first two cases utilize this LB solver independently from the central moment LB methods for interface tracking and two-fluid motions presented in appendices. The first case is that of a static drop inside a periodic box that has a prescribed surfactant gradient along the interface. After some time, the interfacial diffusion process causes the concentration to become more uniformly distributed across the interface. We will compare the spatial distributions and time variations of the concentration field to that of an analytical solution for this case at a variety of dimensionless times and interfacial diffusion coefficients. The second case builds on the first case in that it also includes a prescribed constant background flow field. The domain is extended by double in the direction of the flow, and the drop becomes advected while undergoing the same diffusion process along the interface. Again, we compare the final results with an analytical solution.

Finally, we couple the three central moment-based LB solvers together for the transport of interfacial surfactant concentration, tracking of interfaces, and two-fluid motions in a case of a surfactant-laden drop in surfactant-laden shear flow. The top and bottom horizontal walls of a channel move with constant velocity in opposite directions causing the continuous phase to shear, while the two vertical boundaries are made periodic. Then a drop is placed in the center of the domain and is made to deform by the flow in the bulk region. The surface tension forces tend to cause the drop to revert back to a circular shape while the viscous forces tend to deform the drop further. We modulate this behavior via varying the dimensionless parameter known as the capillary number and compare the deformation parameter to that of an analytical solution in the creeping flow regime with and without the presence of surfactants. Lastly, we demonstrate the fidelity of our approach by simulating the break up of the drop involving inertial effects at higher Reynolds numbers and capillary numbers for cases with and without surfactants. This also serves to demonstrate how the presence of insoluble surfactants on the interface modulate the breakup process of drops under shear.

%%%%%%%%%%%%%%%%%%%%%%%%%%%%%%%%%%%%%%%%%%%%%%%%%%%%% ----------------- BENCHMARK 1
\subsection{Transient interfacial surfactant diffusion on a static drop}
In this test case, a static drop of radius $R$ centered in a periodic box at a location ($x_o, y_o$) is initialized with an angular gradient of surfactant concentration along the interface. After some time, the surfactant is diffused along the surface of the drop such that it tends to move towards less concentrated areas as to become more uniformly distributed along the surface of the drop. The initially imposed nonuniform concentration field is given by
\begin{equation}
\hat{c}_i(\theta) = \frac{1}{2} \left(1 - \cos \theta \right),
\label{ISS_BM1_ana}
\end{equation}
where $\theta$ is the polar angle measured relative the horizontal $x$ axis. In the simulations, this initial condition is imposed via using the equilibrium profile stated in Eq.(\ref{ISSeq}), which then reads as
\begin{equation*}
c = \frac{\hat{c}_i(\theta)}{\cosh^2\left(\frac{2\zeta}{W}\right)},
\end{equation*}
where $\zeta$ is the interface normal location given by $\zeta = R- \sqrt{(x-x_o)^2 + (y-y_o)^2}$.
%\begin{equation*}
%\zeta = R- \sqrt{(x-x_o)^2 + (y-y_o)^2}.
%\end{equation*}
Then, the transient solution for the surfactant concentration along the interface is obtained by solving the partial differential equation of surface diffusion given by $\partial \hat{c}/\partial t = D\sbs{s}\partial^2 \hat{c}/\partial \theta^2$ with the initial condition given in Eq.~(\ref{ISS_BM1_ana}). The resulting analytical solution reads as
\begin{equation}
\hat{c}(\theta,t) = \frac{1}{2} \left(1 - e^{-\frac{D\sbs{s}}{R^2}t}\cos \theta \right),
\end{equation}
where $R\sps{2}/D\sbs{s}$ is the dimensionless time associated with the diffusion process. The results that follow are obtained by using a domain resolved with a grid resolution of $L\sbs{o}\times L\sbs{o}$, where $L\sbs{o}=512$, and the drop diameter $D$ is defined as being half of the domain width as $D= L\sbs{o}/2$. Here, and in the rest of this paper, we report the choice of parameters in terms of the usual lattice units, which is natural for setting up the simulations in LBM. We define our interface width $W$ as $W=4$, and the diffusion coefficient $D\sbs{s}$ set to be $D\sbs{s}=0.1$.

\begin{figure}[H]
\centering
\begin{subfigure}{0.48\textwidth}
\includegraphics[trim = 40 0 40 0, clip, width =70mm]{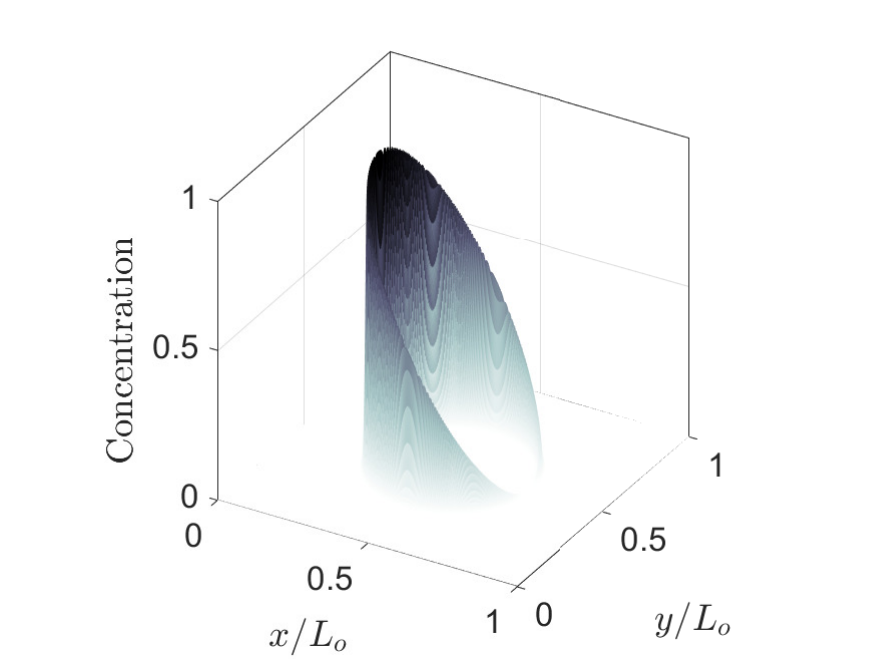}
\caption{$t^*=0$}
\label{ISSBM1c}
\end{subfigure}
\begin{subfigure}{0.48\textwidth}
\includegraphics[trim = 40 0 40 0, clip, width =70mm]{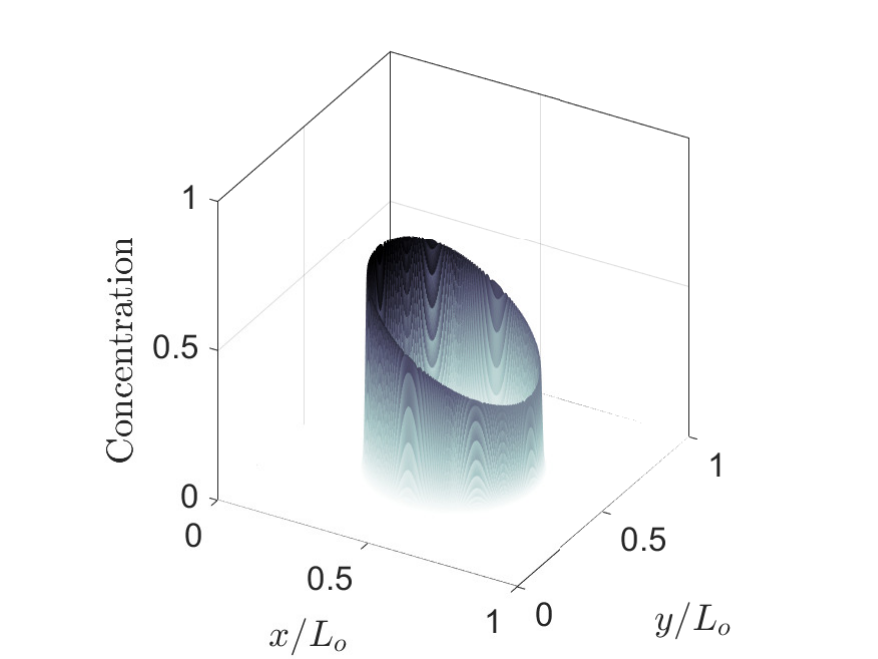}
\caption{$t^*=1$}
\label{ISSBM1d}
\end{subfigure}
\caption{Transient diffusion of an insoluble surfactant concentration field along the interface of a static drop. (a) is the initial imposed concentration field, while (b) is the concentration after unit dimensionless time, i.e., $t\sps{*}=1$.}
\end{figure}

\begin{figure}[H]
\centering
\begin{subfigure}{0.48\textwidth}
\includegraphics[trim = 0 0 0 0, clip, width =70mm]{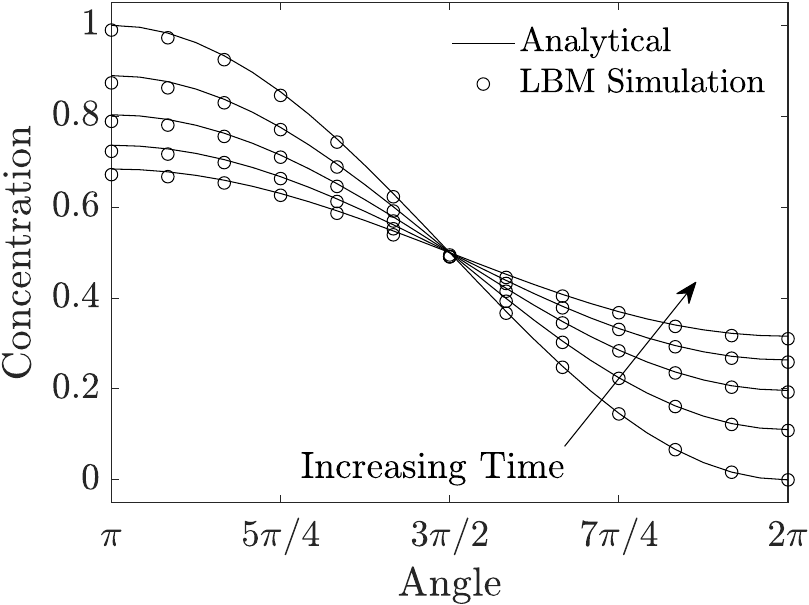}
\caption{}
\label{timecomp}
\end{subfigure}
\begin{subfigure}{0.48\textwidth}
\includegraphics[trim = 0 0 0 0, clip, width =70mm]{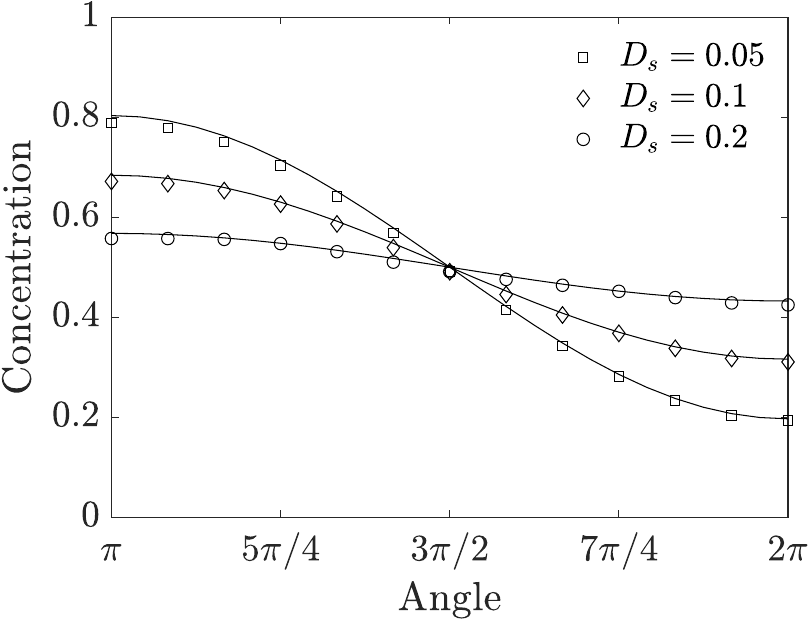}
\caption{}
\label{diffusion}
\end{subfigure}
\caption{Profiles of insoluble surfactant concentration along the interface of a static drop that undergo variations due to surface diffusion. (a) Comparison of the computed results using the central moment LBM with the analytical solution at different dimensionless times $t^*$  = (0, 0.25, 0.5, 0.75, 1.0). (b) with different values of the diffusion coefficient $D_s$ = (0.05, 0.1, 0.2) presented at the same dimensionless time $t^*=1$.}
\end{figure}

Figure~\ref{ISSBM1c} shows the magnitude of the initial concentration while Fig.~\ref{ISSBM1d} shows the surfactant concentration magnitude after one dimensionless time step $t\sps{*}$ where $t\sps{*}=t/T\sbs{D}$ with $T\sbs{D}=R\sps{2}/D\sbs{s}$. These figures clearly indicate how the surfactant is becoming diffused along the interfacial surface of the drop.
Moreover, in Fig.~\ref{timecomp}, we show a direct quantitative comparison of the simulation results obtained using the central moment LBM (symbols) with the analytically predicted solution (curves) for different angles $\theta$ along the surface of the drop and at a variety of time stamps. Good agreement is found for all the times shown. In addition, Fig.~\ref{diffusion} shows a comparison with different diffusion coefficients, $D\sbs{s}$ at a dimensionless time of $t^*=1$. Again, good general agreement is observed. Thus, our central moment-based LB scheme for evolving interface-bound scalar field is shown to yield accurate solutions. The method is free of any singularity issue and is able to confine the concentration field exclusively on the interface quite well.

%%%%%%%%%%%%%%%%%%%%%%%%%%%%%%%%%%%%%%%%%%%%%%%%%%%%% ----------------- BENCHMARK 2
\subsection{Transient interfacial surfactant diffusion on an uniformly advected drop}
\begin{figure}[H]
\centering
\begin{subfigure}{0.68\textwidth}
\includegraphics[trim = 0 0 0 0, clip, width =100mm]{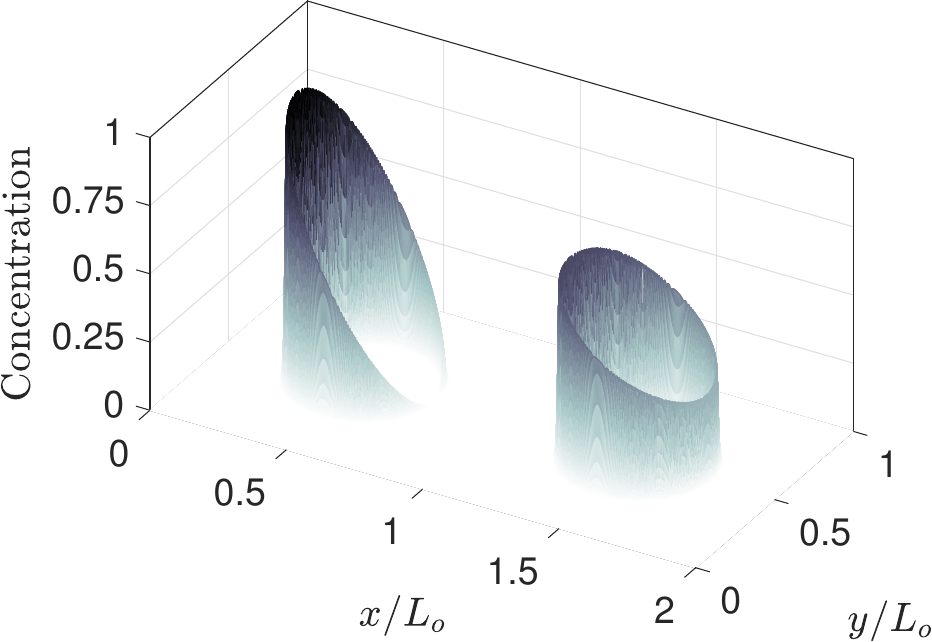}
\end{subfigure}
\caption{Diffusion of an insoluble surfactant concentration along the interface of a drop advected in a constant and uniform horizontal flow field in the $x$-direction with magnitude $U_o$ at times $t^*=0$ and $t^*=1$.}
\label{ISS_BM2_2}
\end{figure}

In this next test case, we repeat of the first case but with a prescribed background velocity of constant and uniform magnitude, $U\sbs{o}$. The velocity is chosen such that the drop will travel exactly a distance of $L\sbs{o}$ in one dimensionless time unit $t\sps{*}=t/T\sbs{D}=1$ with $T\sbs{D}=R\sps{2}/D\sbs{s}$. In this regard, the domain is extended in the flow direction such that it is twice as long as it is wide, i.e., we resolve the domain with $2L\sbs{o}\times L\sbs{o}$ with grid nodes, where $L\sbs{o}=512$. All the boundaries are made periodic. Again, we let the diameter of the drop be half of the domain height, $D = L\sbs{o}/2$, the diffuse interface width is set to $W=4$, and the diffusion coefficient is $D\sbs{s}=0.1$. In this case, we use Eq.~(\ref{ISS_BM1_ana}) as the reference analytical solution for comparison, but at a shifted location $(x\sbs{o} + U\sbs{o}\Delta t,y\sbs{o})$ after a dimensionless time of $t^*=1$. Figure~\ref{ISS_BM2_2} presents the simulation results at the initial condition (left) and after one dimensionless time unit (right). Good agreement between the simulation results and the analytical prediction for the surfactant concentration field even as the drop advects uniformly is observed.

%%%%%%%%%%%%%%%%%%%%%%%%%%%%%%%%%%%%%%%%%%%%%%%%%%%%% ----------------- BENCHMARK 3
\subsection{Deformation and break-up processes of insoluble surfactant-laden drop in shear flow}

\begin{figure}[H]
\centering
\begin{subfigure}{0.48\textwidth}
\includegraphics[trim = 0 0 0 0, clip, width =65mm]{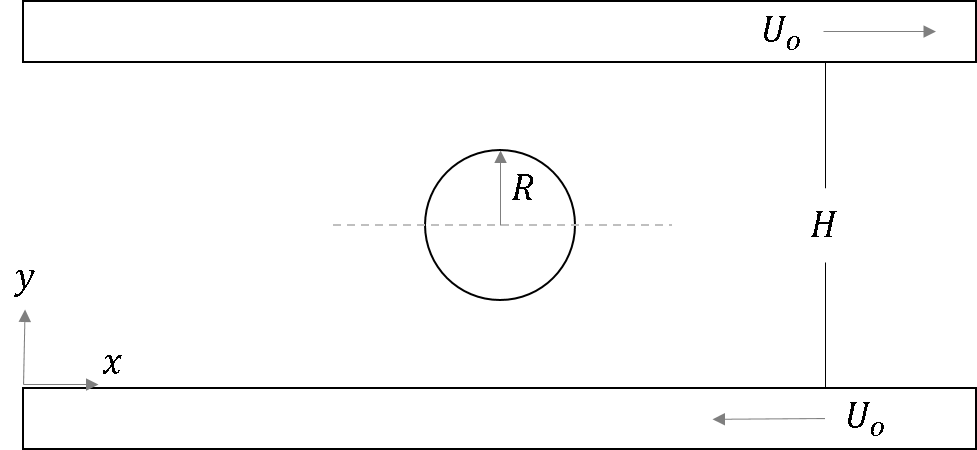}
\end{subfigure}
\caption{Schematic of the insoluble surfactant-laden drop in shear flow problem set up.}
\label{SF}
\end{figure}

\begin{figure}[H]
\centering
\begin{subfigure}{0.25\textwidth}
\includegraphics[trim = 0 0 0 0, clip, width =25mm]{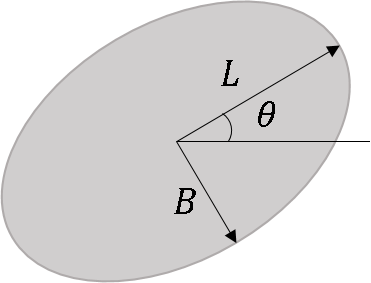}
\end{subfigure}
\caption{Schematic description of the drop deformation, where the Taylor deformation parameter $D$ is computed via $D =(L-B) / (L+B)$.}
\label{drop_deformation}
\end{figure}

This final case study consists of an insoluble surfactant-laden drop placed in a shear flow within a channel whereby the upper and lower plates move in opposite directions with the same speed, $U\sbs{o}$, while the two vertical boundaries on each side are considered periodic. The flow domain is resolved with $3H\times H$ grid nodes, where $H$ is the channel height which is set $H=256$; the drop has a radius one eighth of the channel height, i.e., $R = H/8$. A graphical description of this problem set up is illustrated in Figure~\ref{SF} for convenience.
\noindent
The flow field without a drop would reach a steady state solution of the following form, and here we use this as an initial condition
\begin{equation*}
u(y) = U\sbs{o}\left( 2 \frac{y}{H} - 1 \right).
\end{equation*}
\noindent
Then, a drop is placed at the center of the domain filled with a neutrally buoyant fluid (i.e., with a density ratio of unity) and experiences shear from the upper and lower parts of the flow field in opposing directions, and thus deforms anti-symmetrically across the domain half channel. Then, the magnitude of the shear rate experienced by the fuids is $\dot{\gamma} = 2 U\sbs{o}/H$. The primary dimensionless number that characterizes the drop deformations is the capillary number $\mbox{Ca} =\mu \dot{\gamma}r/\sigma\sbs{o}$, which is used to modulate the magnitude of the viscous forces relative to the surface tension forces. These forces tend to compete in the drop deformation process as the viscous drag causes drop deformation while the surface tension tends to cause the deformed drop to revert back into a circular shape. Moreover, the Reynolds number is also used here to modulate the inertial forces relative to the viscous forces and is defined as $\mbox{Re}=\dot{\gamma}R\sps{2}/\nu$. Here, we let both the density ratio and viscosity ratio set to one for simplicity, and use $\rho\sbs{a}=\rho\sbs{b} = 1$ and  $\nu\sbs{a}=\nu\sbs{b} = 0.1$. In the following, from results obtained in our simulations, we compare the dimensionless drop deformation parameter $D$, which is defined and illustrated in Fig.~\ref{drop_deformation},
\noindent
with an analytical solution for insoluble surfactant-laden drop in the creeping flow regime derived by Stone and Leal~\cite{stone1990effects}. Here, the analytically predicted deformation parameter $D$ is given by~\cite{stone1990effects}
\begin{equation}
D \approx \frac{3 \mbox{Ca} \; b\sbs{r} }{4 + \mbox{Ca} \; b\sbs{r}},
\end{equation}
\noindent
where
\begin{equation}
b\sbs{r} = \frac{5}{4} \frac{(16+19\lambda) + 4\beta\left(\frac{\mbox{Pe}}{\mbox{Ca}}\right)}{10(1+\lambda) + 2\beta\left(\frac{\mbox{Pe}}{\mbox{Ca}}\right)}.
\end{equation}
\noindent
Here, $\beta$ is the Gibbs elasticity parameter, which measures the sensitivity of the surface tension to the surfactant concentration (with $\beta=0$ referring to the clean interface case), $\mbox{Pe}$ is the Peclet number given by $\mbox{Pe}=\mbox{Re}(\nu/D\sbs{s})$, and $\lambda$ is the viscosity ratio.

\begin{figure}[H]
\centering
\begin{subfigure}{0.5\textwidth}
\includegraphics[trim = 0 0 0 0, clip, width =75mm]{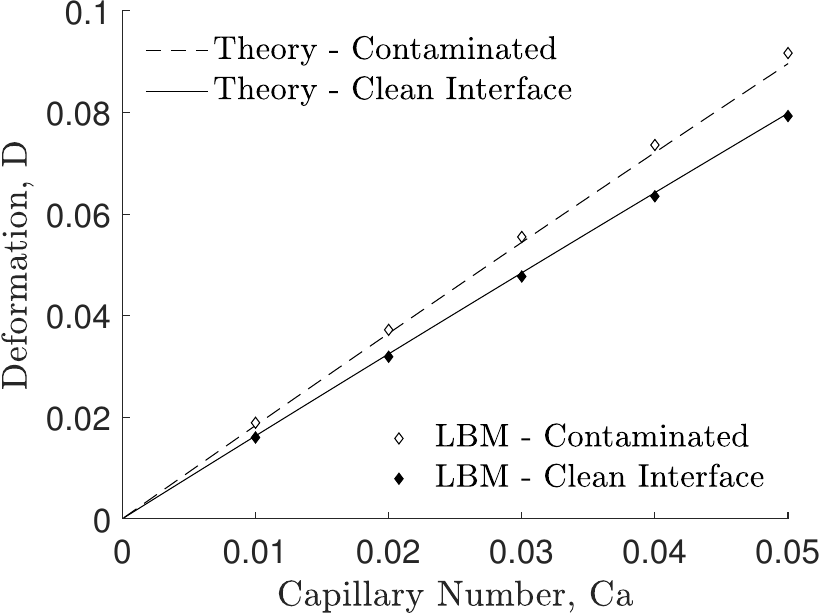}
\end{subfigure}
\caption{A comparison of the deformation parameters $D$ at various capillary numbers $\mbox{Ca}$ for clean ($\beta = 0$) and contaminated ($\beta = 0.25$) interfaces between the simulation results computed using the central moment LBM at a Reynolds number $\mbox{Re}=0.1$ (symbols) with the analytically predicted values based on the creeping flow theory of Stone and Leal (lines)~\cite{stone1990effects}.}
\label{SD_Analytical}
\end{figure}

To begin with, we compare simulation results with this prediction across a range of Capillary numbers, and we set Reynolds number equal to 0.1. Figure~\ref{SD_Analytical} shows a direct comparison between our LBM simulation results and the above analytical solution for the clean ($\beta=0$) and surfactant-laden or contaminated ($\beta=0.25$) surface of the drop under shear with the former numerical data listed further in Table~\ref{SD_Table}. In general, as the local surface tension decreases due to the presence of surfactant concentration field in the contaminated case, it results in greater deformation relative to the clean interface case; furthermore, as $\mbox{Ca}$ increases, the drop experiences greater deformation. These qualitative trends are reproduced with our central moment LBM based simulations, which are also in good quantitative agreement with theory of Stone and Leal~\cite{stone1990effects}

\begin{table}[H]
\centering
\begin{tabular}{c c c}
\toprule
Capillary Number & &  Deformation Parameter\\
$\mbox{Ca}$  & &   $\qquad\qquad\qquad D\sbs{\mbox{clean}}$\qquad $D\sbs{\mbox{contaminated}}$ ($\times10\sps{-2}$) \\
%\hline
\midrule
0.01  & &  \quad 1.60\qquad\qquad\quad 1.89   \\
0.02  & &  \quad 3.19\qquad\qquad\quad 3.72   \\
0.03  & &  \quad 4.77\qquad\qquad\quad 5.55    \\
0.04  & &  \quad 6.35\qquad\qquad\quad 7.36   \\
0.05  & &  \quad 7.93\qquad\qquad\quad 9.17    \\
\bottomrule
\end{tabular}
\caption{Listing of the numerical simulation results computed using central moment LBM (shown as symbols in Fig.~\ref{SD_Analytical}) for the deformation parameter $D$ for clean and contaminated (i.e., surfactant-laden) interfaces for different capillary numbers $\mbox{Ca}$ at $\mbox{Re}=0.1$.}
\label{SD_Table}
\end{table}
We mention here that the results in this case are sensitive to the mobility parameter in the conservative Allen Cahn equation that governs the diffusion and sharpening of the order parameter. Xu et al.~\cite{xu2023high} pointed out that to obtain a balance between the boundedness of the order parameter and the accuracy, a compromise must be made such that the Peclet number based on the interface width, $Pe\sps{*}=U\sbs{o}W/M\sbs{\phi}$ be on the order of one, $O(1)$. This is consistent with the asymptotic analysis performed earlier by Magaletti et al.~\cite{magaletti2013sharp}. In this work, we let $\mbox{Pe}\sps{*}=0.5$.

\begin{figure}[H]
\centering
\begin{subfigure}{0.488\textwidth}
\includegraphics[trim = 0 0 0 0, clip, width =72mm]{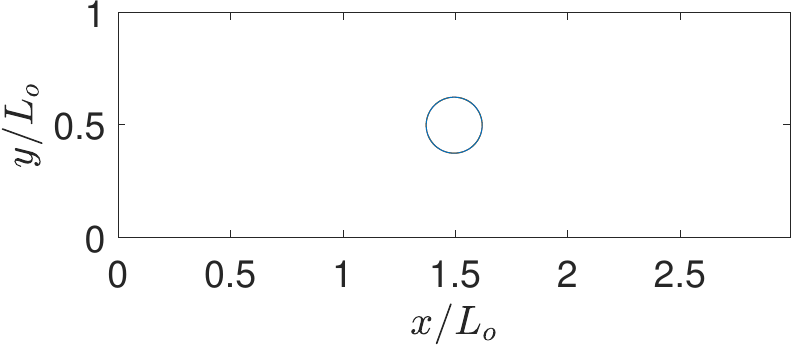}
\caption{$t^* =0$}
\end{subfigure}
\begin{subfigure}{0.48\textwidth}
\includegraphics[trim = 0 0 0 0, clip, width =72mm]{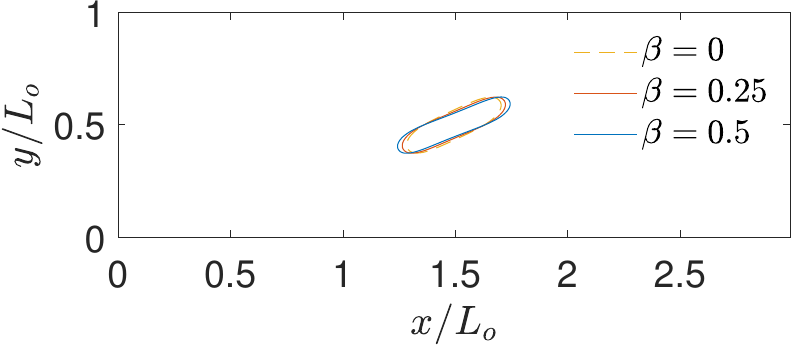}
\caption{$t^* =5$}
\end{subfigure}

\begin{subfigure}{0.488\textwidth}
\includegraphics[trim = 0 0 0 0, clip, width =72mm]{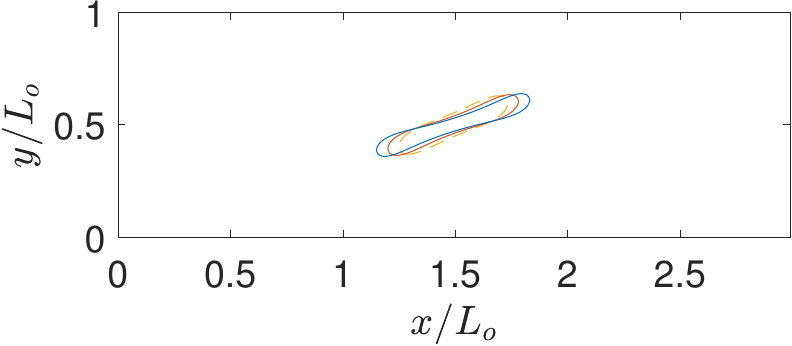}
\caption{$t^* =10$}
\end{subfigure}
\begin{subfigure}{0.48\textwidth}
\includegraphics[trim = 0 0 0 0, clip, width =72mm]{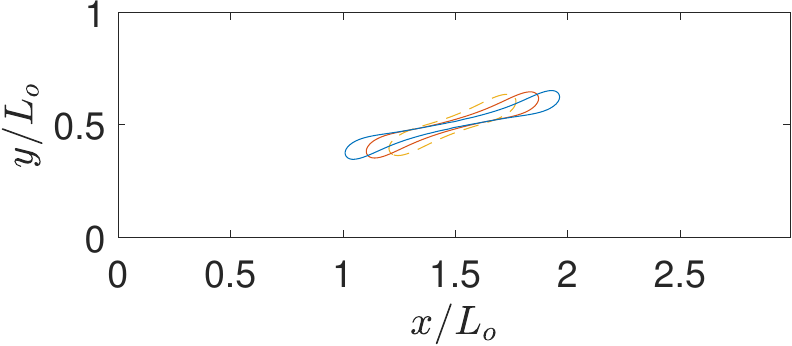}
\caption{$t^* = 15$}
\end{subfigure}

\begin{subfigure}{0.488\textwidth}
\includegraphics[trim = 0 0 0 0, clip, width =72mm]{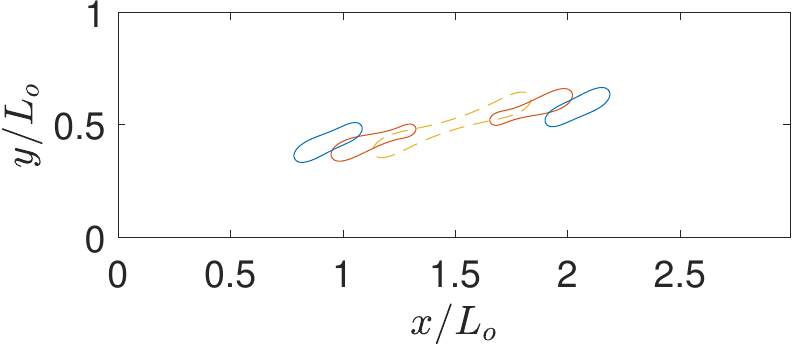}
\caption{$t^* =20$}
\end{subfigure}
\begin{subfigure}{0.48\textwidth}
\includegraphics[trim = 0 0 0 0, clip, width =72mm]{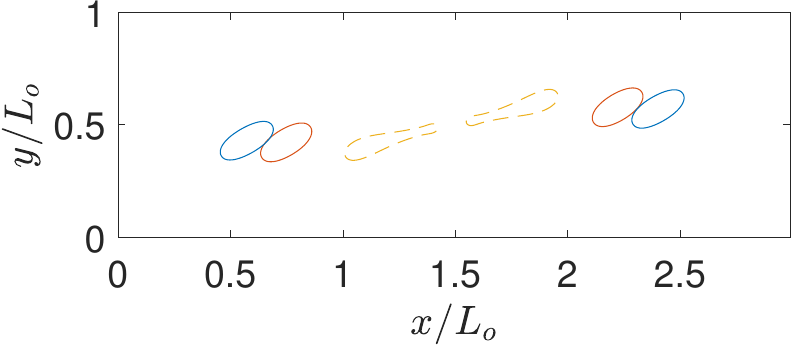}
\caption{$t^* = 25$}
\end{subfigure}

\caption{Time sequence of a drop in shear flow with and without insoluble surfactants at $\mbox{Re}=1$ and $\mbox{Ca}=0.35$ with $\beta=0$ (clean interface) and $\beta=0.25$ and 0.50 (surfactant-laden interface). The dimensionless time $t^*$ is based on the deformation rate to define the reference scale so that $t^*=\dot{\gamma}t$. }\label{fig:breakupRe1}
\end{figure}

Next, we demonstrate the ability of our novel LB approach in simulating the dynamics of drop break-up processes, which occur with higher inertial effects and/or viscous forces relative to surface tension than before, or, equivalently, at higher Reynolds and capillary numbers by considering cases with ($\beta\neq 0$) and without ($\beta=0$) surfactant on interfaces. Instead of attaining a steady-state of the deformed state for the drop as seen earlier at low $\mbox{Re}\ll 1$ and relatively small $\mbox{Ca}$, the viscous and inertial forces dominate over the surface tension at higher $\mbox{Re}$ and $\mbox{Ca}$ such that the drop undergoes continuous deformation and stretching and ultimately breaking up. We will also show that the presence of surfactant on the interface of a shear-driven drop can breakup more readily when compared to the clean interface case. Figure~\ref{fig:breakupRe1} presents simulation results at $\mbox{Re}=1$ and $\mbox{Ca}=0.35$, which show time sequence of the dynamics of the sheared drop at various dimensionless times $t^* = \dot{\gamma}t$ for $\beta=0$ (clean interface) and $\beta=0.25$ and 0.50 (surfactant-laden interface). It can be seen that in all these cases, the drop eventually breaks-up, which is more prominently facilitated by the presence of surfactant with higher sensitivity to surface tension or $\beta$; by contrast, the clean interface case results in a break-up of drop at a later stage than the surfactant-laden case. Moreover, the deformation process and the break-up patterns of the drop are significantly different as $\beta$ is increased. We find that the dimensionless drop breakup times for the parameters chosen are consistent with other simulation results found in the literature.
We also present simulation results from another case, where we increase the Reynolds number further to $\mbox{Re}=4$. Under greater shearing rates in such cases, the drop dramatically stretches out almost horizontally in the form of relatively thin filaments; this can then also take on additional modes and breaks apart into multiple drops instead of two. This behavior is qualitatively consistent with the results presented by Wagner et al.~\cite{wagner2003role}. The presence of surface acting agents acts to exaggerate this behavior further. Moreover, from these simulations, it seems that the presence of surfactant has greater effect on the drop deformation process at lower $\mbox{Re}$.
\begin{figure}[H]
\centering
\begin{subfigure}{0.488\textwidth}
\includegraphics[trim = 0 0 0 0, clip, width =72mm]{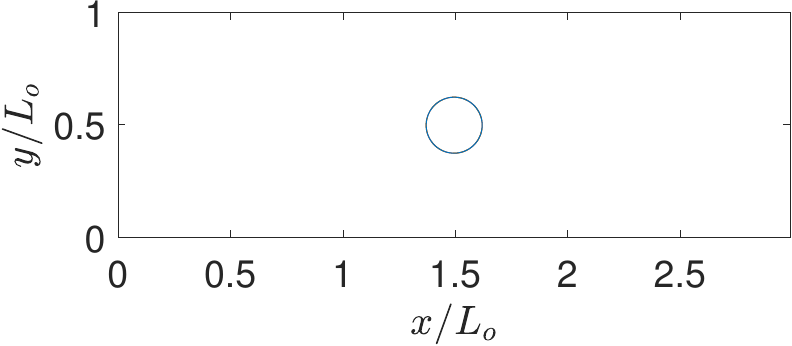}
\caption{$t^* =0$}
\end{subfigure}
\begin{subfigure}{0.48\textwidth}
\includegraphics[trim = 0 0 0 0, clip, width =72mm]{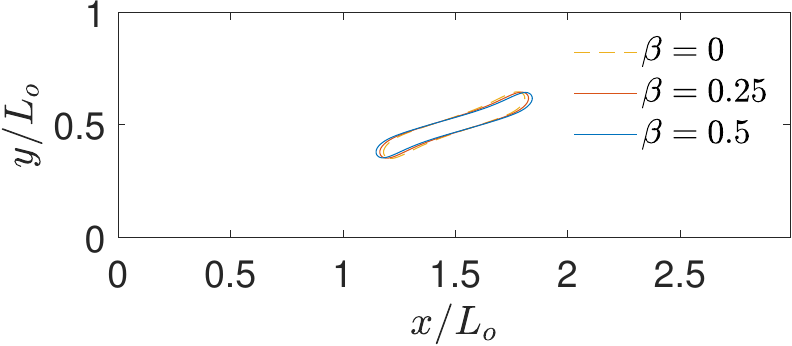}
\caption{$t^* =5$}
\end{subfigure}

\begin{subfigure}{0.488\textwidth}
\includegraphics[trim = 0 0 0 0, clip, width =72mm]{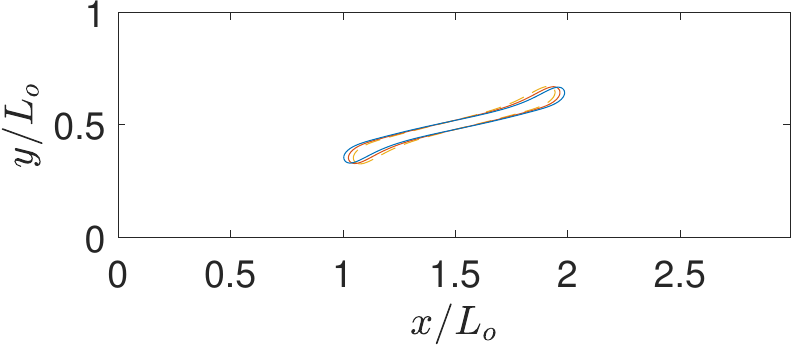}
\caption{$t^* =7.5$}
\end{subfigure}
\begin{subfigure}{0.488\textwidth}
\includegraphics[trim =0 0 0 0, clip, width =72mm]{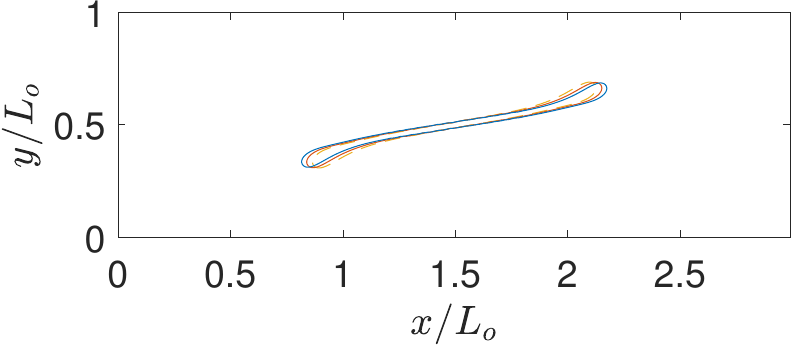}
\caption{$t^* =10$}
\end{subfigure}

\begin{subfigure}{0.488\textwidth}
\includegraphics[trim =0 0 0 0, clip, width =72mm]{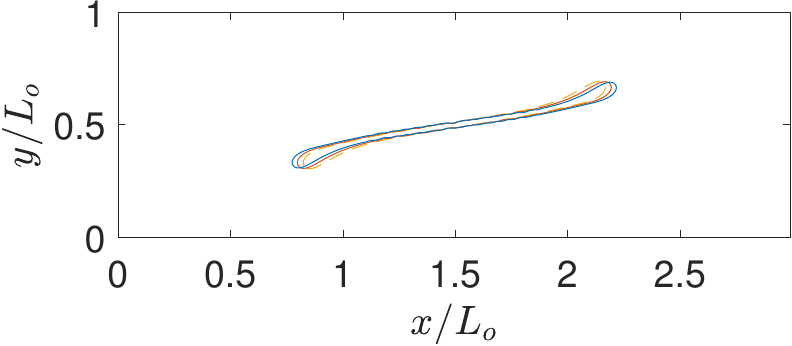}
\caption{$t^* =10.5$}
\end{subfigure}
\begin{subfigure}{0.488\textwidth}
\includegraphics[trim =0 0 0 0, clip, width =72mm]{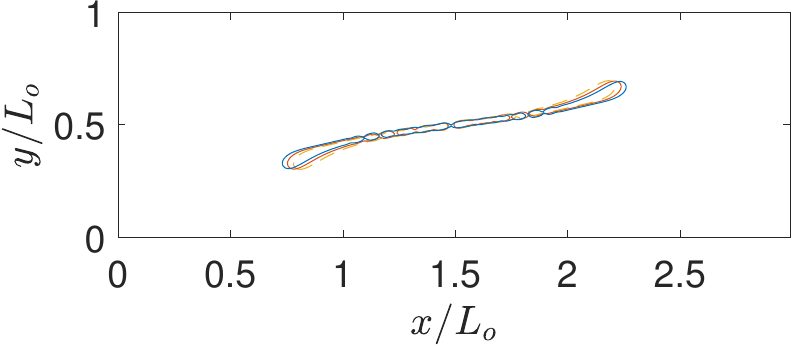}
\caption{$t^* =11$}
\end{subfigure}
\caption{Time sequence of a drop in shear flow with and without insoluble surfactants at $\mbox{Re}=4$ (which is four times higher than that in Fig.~\ref{fig:breakupRe1}) and $\mbox{Ca}=0.35$ with  $\beta=0$ (clean interface) and $\beta=0.25$ and 0.50 (surfactant-laden interface). The dimensionless time $t^*$ is based on the deformation rate to define the reference scale so that $t^*=\dot{\gamma}t$.}\label{fig:breakupRe4}
\end{figure}

%%%%%%%%%%%%%%%%%%%%%%%%%%%%%%%%%%%%%%%%%%%%%%%%%%%%% ----------------- CONCLUSION
\section{Summary and Conclusions} \label{sec:summaryandconclusions}
In this work, we present a new LB method using a robust central moment-based collision model to solve a recently derived formulation for the transport of an interface-confined scalar field. The latter is modeled as a regularized advection-diffusion equation on a diffuse interface, which has applications related to insoluble surfactant-laden multiphase flows. We then demonstrated the fidelity of this novel LB approach in a series of benchmark test cases. The first case demonstrated the diffusion of the surfactant concentration with a nonuniform initial distribution along the interface of a static drop is accurate in that quantitative comparisons of the computed concentration profiles at various times with an analytical solution were found to match quite well. The second test case involved an imposed background flow field that does not modify the shape of the same drop but causes the drop to become advected and the temporal variation of surfactant diffusion on the interface in such a case was again consistent with the analytical solution.

Lastly, the third case considered the simulations of surfactant-laden drop in shear flow, which required coupling the central moment-based LB scheme for interface-confined surfactant field with two additional LB solvers, which are again based on central moments, for interface and two-fluid motions in a triple distribution functions-based approach. In this regard, we use the Langmuir isotherm as the interface equation of state to couple the surfactant concentration gradients to the fluid motions via normal and tangential surface tension forces. At low Reynolds numbers, i.e., in the creeping flow regime, the computed Taylor deformation parameter of the drop at steady state at various capillary numbers was found to be in good quantitative agreement with the theory due to Stone and Leal for both clean and contaminated interface cases, with the latter exhibiting larger deformation than the former. Lastly, we showed that with larger Reynolds numbers and capillary numbers, the drop breaks up under higher inertial and viscous forces and the onset of this fragmentation was found to be consistent with those found in literature for the same parameters. Moreover, the presence of surfactants was found to facilitate the break up process. The central moment LBM for interface-confined transport of concentration field is thus accurate, robust and free of any singularity issues in phase field modeling and simulation of surfactant-laden multiphase flows and its extension to three-dimensions is straightforward. Moreover, the approach can be further adapted for soluble surfactants which is a subject for future investigations.

\section*{Acknowledgements}
Investigations on this research were presented at 32nd International Conference on Discrete Simulation of Fluid Dynamics (DSFD), Albuquerque, New Mexico, July 2023 and at American Physical Society (APS) 76th Annual Division of Fluid Dynamics (DFD) Meeting, Washington D.C., November 2023~(https://meetings.aps.org/Meeting/DFD23/Session/T16.6). The authors would like to acknowledge the support of the US National Science Foundation (NSF) for research under Grant CBET-1705630 and also thank the NSF for support of the development of a computer cluster `Alderaan' hosted at the Center for Computational Mathematics at the University of Colorado Denver under Grant OAC-2019089 (Project ``CC* Compute: Accelerating Science and Education by Campus and Grid Computing''), which was used in performing the simulations.

\section*{Data Availability Statement}
Data generated during the current study are available from the corresponding author on reasonable request.

\appendix

\section{Distribution Functions-to-Raw Moments Transformation Matrix and Inverse}\label{A}
For the D2Q9 lattice, the transformation matrix $\tensr{P}$ is used to transform the discrete distribution functions into corresponding raw moments, which can be explicitly written as
\begin{equation}
\tensr{P} = \begin{bmatrix}
     1  &\quad    1  &\quad    1  &\quad      1  &\quad    1  &\quad     1 &\quad    1  &\quad     1  &\quad     1 \\[10pt]
     0  &\quad    1  &\quad     0  &\quad    \um1  &\quad     0  &\quad    1 &\quad   \um1 &\quad   \um1  &\quad     1 \\[10pt]
     0  &\quad    0  &\quad     1  &\quad     0  &\quad   \um1  &\quad    1  &\quad     1  &\quad    \um1  &\quad  \um1 \\[10pt]
     0  &\quad    1  &\quad     0  &\quad     1  &\quad     0  &\quad    1  &\quad     1  &\quad     1  &\quad     1 \\[10pt]
     0  &\quad    0  &\quad     1  &\quad     0  &\quad     1  &\quad    1  & \quad    1  &\quad     1  &\quad     1 \\[10pt]
     0  &\quad    0  &\quad     0  &\quad     0  &\quad     0  &\quad    1  &\quad    \um1  &\quad     1  &\quad   \um1 \\[10pt]
     0  &\quad    0  &\quad     0  &\quad     0  &\quad     0  &\quad    1  &\quad     1  &\quad  \um1  &\quad  \um1 \\[10pt]
     0  &\quad    0  &\quad     0  &\quad     0  &\quad     0  &\quad    1  &\quad    \um1  &\quad  \um1  &\quad    1 \\[10pt]
     0  &\quad    0  &\quad     0  & \quad    0  &\quad     0  & \quad   1  &\quad     1  &\quad     1  &\quad    1
\end{bmatrix}
\label{Pmatrix}
\end{equation}
The inverse of the transformation matrix $\tensr{P}$ is then used to map from nine independent raw moments to equivalent distribution functions and the $\tensr{P}^{-1}$ reads as
\begin{equation}
\tensr{P}^{-1} =
\begin{bmatrix}
     1  &\quad    0  &\quad    0  &\quad    \um 1  &\quad   \um 1  &\quad     0 &\quad    0  &\quad     0  &\quad     1 \\[10pt]
     0  &\quad    \frac{1}{2}  &\quad     0  &\quad    \frac{1}{2}  &\quad     0  &\quad    0 &\quad   0 &\quad   \um \frac{1}{2}  &\quad     \um \frac{1}{2} \\[10pt]
     0  &\quad    0  &\quad     \frac{1}{2}  &\quad     0  &\quad   \frac{1}{2}  &\quad    0  &\quad     \um \frac{1}{2}  &    0  &\quad   \um \frac{1}{2} \\[10pt]
     0  &\quad    \um \frac{1}{2}  &\quad     0  &\quad     \frac{1}{2}  &\quad     0  &\quad   0 &\quad     0  &\quad     \frac{1}{2}  &\quad    \um \frac{1}{2} \\[10pt]
     0  &\quad    0  &\quad     \um \frac{1}{2}  &\quad     0  &\quad     \frac{1}{2}  &\quad    0  &\quad     \frac{1}{2} &     0  &\quad     \um \frac{1}{2} \\[10pt]
     0  &\quad    0  &\quad     0  &\quad     0  &\quad     0  &\quad    \frac{1}{4}  &\quad    \frac{1}{4}  &\quad     \frac{1}{4}  &\quad   \frac{1}{4} \\[10pt]
     0  &\quad    0  &\quad     0  &\quad     0  &\quad     0  &\quad    \um \frac{1}{4}  &\quad     \frac{1}{4}  &\quad  \um \frac{1}{4}  &\quad \frac{1}{4} \\[10pt]
     0  &\quad    0  &\quad     0  &\quad     0  &\quad     0  &\quad    \frac{1}{4}  &\quad    \um \frac{1}{4} &\quad  \um \frac{1}{4}  &\quad    \frac{1}{4} \\[10pt]
     0  &\quad    0  &\quad     0  &\quad     0  &\quad     0  &\quad    \um \frac{1}{4}  &\quad     \um \frac{1}{4}  &\quad     \frac{1}{4}  &\quad    \frac{1}{4}
\end{bmatrix}
\label{Pinv}
\end{equation}

\section{Raw Moments-to-Central Moments Transformation Matrix and Inverse}\label{B}
The transformation matrix $\tensr{F}$ is used to map the nine independent raw moments of the distribution functions into the respective central moments for the D2Q9 lattice. It can be represented as follows:
\begin{equation}
\tensr{F}=
\begin{bmatrix}
      1  &    0  &    0  &     0  &    0  &     0 &    0  &     0  &     0 \\[10pt]

     \um u_x  &   1  &    0  &     0  &    0  &     0 &    0  &     0  &     0 \\[10pt]

      \um u_y  &    0  &   1  &     0  &    0  &     0 &    0  &     0  &     0 \\[10pt]

      u_x^2  &   \um 2u_x  &    0  &     1  &    0  &     0 &    0  &     0  &     0 \\[10pt]

      u_y^2  &    0  &    \um 2u_y  &     0  &    1  &     0 &    0  &     0  &     0 \\[10pt]

     u_x u_y  &   \um u_y  &    \um u_x  &     0  &    0  &     1 &    0  &     0  &     0 \\[10pt]

      \um u_x^2 u_y  &   2u_x u_y  &   u_x^2  &     \um u_y  &    0  &     \um 2u_x &    1  &     0  &     0 \\[10pt]

      \um u_x u_y^2 &    u_y^2  &    2u_x u_y  &     0  &    \um u_x  &    \um 2u_y &    0  &     1  &     0 \\[10pt]

      u_x^2 u_y^2  &    \um u_x u_y^2  &    \um u_x^2 u_y  &    u_y^2  &   u_x^2  &    4u_x u_y &    \um 2 u_y  &     \um 2  u_x  &     1 \\
\end{bmatrix}
\label{F}
\end{equation}
The inverse transformation matrix $\tensr{F}^{-1}$ is then used to transform from nine central moments into the respective raw moments, and is given as
\begin{equation}
\tensr{F}^{-1}=
\begin{bmatrix}
      1  &    0  &    0  &     0  &    0  &     0 &    0  &     0  &     0 \\[10pt]

      u_x  &   1  &    0  &     0  &    0  &     0 &    0  &     0  &     0 \\[10pt]

      u_y  &    0  &   1  &     0  &    0  &     0 &    0  &     0  &     0 \\[10pt]

      u_x^2  &   2u_x  &    0  &     1  &    0  &     0 &    0  &     0  &     0 \\[10pt]

      u_y^2  &    0  &    2u_y  &     0  &    1  &     0 &    0  &     0  &     0 \\[10pt]

     u_x u_y  &   u_y  &    u_x  &     0  &    0  &     1 &    0  &     0  &     0 \\[10pt]

      u_x^2 u_y  &   2u_x u_y  &   u_x^2  &     u_y  &    0  &     2u_x &    1  &     0  &     0 \\[10pt]

      u_x u_y^2 &    u_y^2  &    2u_x u_y  &     0  &    u_x  &    2u_y &    0  &     1  &     0 \\[10pt]

      u_x^2 u_y^2  &    u_x u_y^2  &    u_x^2 u_y  &    u_y^2  &   u_x^2  &    4u_x u_y &    2 u_y  &     2  u_x  &     1 \\
\end{bmatrix}
\label{Finv}
\end{equation}

\section{Central moment LBM for Interface Tracking Based on Conservative Allen-Cahn Equation}\label{app:CACE}
The discrete distribution functions $f\sbs{\alpha}$, where $\alpha = 0, 1, \ldots, 8$ are evolved to compute the order parameter used in the conservative Allen-Cahn equation (see Eq.~(\ref{CACE})) for the tracking of interfaces. In order to construct a central moment LBM for $f\sbs{\alpha}$, we first define their discrete raw moments and central moments, respectively, of order ($m+n$)as
\begin{subequations}
\begin{equation}
\qquad \left( \begin{array}{c}\kappa'\sbs{mn}\\[2mm]   \kappa'^{\;\sss{eq}}\sbs{mn} \end{array} \right)  = \sum\sbs{\alpha = 0}\sps{8} \left( \begin{array}{c}f\sbs{\alpha} \\[2mm]   f\sbs{\alpha}\sps{eq} \end{array} \right)  e\sbs{\alpha x}\sps{m}   e\sbs{\alpha y}\sps{n},
\label{rm}
\end{equation}
\begin{equation}
\qquad \left( \begin{array}{c}\kappa\sbs{mn} \\[2mm]   \kappa\sbs{mn}\sps{eq} \end{array} \right)  = \sum\sbs{\alpha = 0}\sps{8} \left( \begin{array}{c}f\sbs{\alpha} \\[2mm]  f\sbs{\alpha}\sps{eq} \end{array} \right) (e\sbs{\alpha x}-u\sbs{x})\sps{m}  ( e\sbs{\alpha y}-u\sbs{y})\sps{n}.
\label{cm}
\end{equation}
\end{subequations}
Their nine independent components for the D2Q9 lattice are represented by the vectors $\bm{\kappa'}$ and $\bm{\kappa}$, respectively, as
\begin{subequations}
\begin{eqnarray}
\qquad \bm{\kappa'} \! \! \! &=& \! \! \! ( \kappa'\sbs{00}, \kappa'\sbs{10},\kappa'\sbs{01}, \kappa'\sbs{20}, \kappa'\sbs{02}, \kappa'\sbs{11},\kappa'\sbs{21}, \kappa'\sbs{12},\kappa'\sbs{22} )\sps{\top},\label{eqn:4a} \\[3mm]
\qquad \bm{\kappa} \! \! \! &=& \! \! \! ( \kappa\sbs{00},\kappa\sbs{10}, \kappa\sbs{01}, \kappa\sbs{20}, \kappa\sbs{02}, \kappa\sbs{11}, \kappa\sbs{21}, \kappa\sbs{12}, \kappa\sbs{22} )\sps{\top}.
\end{eqnarray}
\end{subequations}
Then, to construct the collision step, following~\cite{hajabdollahi2021central}, we determine the discrete central moment equilibria  via a matching principle by matching the discrete equilibrium central moments to the corresponding continuous central moments of the Maxwell distribution by replacing the density $\rho$ with the order parameter $\phi$. Moreover, the interfacial sharpening flux term $M\sbs{\phi}\theta\bm{n}$ in Eq.~(\ref{CACE}) are accounted for via the extended moment equilibria of the first order moments so as to recover the conservative Allen-Cahn equation.
\begin{gather}
\qquad \kappa\sbs{00}\sps{eq} = \phi, \qquad
\kappa\sbs{10}\sps{eq} = M\sbs{\phi} \theta  n\sbs{x},\qquad
\kappa\sbs{01}\sps{eq} = M\sbs{\phi} \theta  n\sbs{y},\nonumber \\[2mm]
\qquad \kappa\sbs{20}\sps{eq} = c\sbs{s}\sps{2}\cdot \phi,\qquad
\kappa\sbs{02}\sps{eq} = c\sbs{s}\sps{2}\cdot \phi,\qquad
\kappa\sbs{11}\sps{eq} = 0,\nonumber  \\[2mm]
\qquad \chi\sbs{21}\sps{eq} = 0,\qquad
\kappa\sbs{12}\sps{eq} = 0,\qquad
\kappa\sbs{22}\sps{eq} = c\sbs{s}\sps{4}\cdot \phi,
\end{gather}

Based on these considerations, we can now write central moment LB equation for interface tracking as
\begin{equation}
f\sbs{\alpha}(\bm{x}+\bm{e}\sbs{\alpha}\Delta t, t+\Delta t) - f\sbs{\alpha}(\bm{x}, t) = \tensr{P}\sps{\um 1}\tensr{F}\sps{\um 1} \left\{ \bm{\Lambda\sbs{\kappa}} \left[ \bm{\kappa}\sps{eq} - \bm{\kappa}\right] \right\},
\label{LBE_CACE}
\end{equation}
where $\bm{\Lambda\sbs{\kappa}}$ is a diagonal relaxation rate matrix is given by $\bm{\Lambda\sbs{\kappa}} = \mbox{diag}(1, \omega\sbs{\kappa}, \omega\sbs{\kappa},1 ,1,1,1,1,1)$. Note that the first order moments are related to the mobility coefficient $M\sbs{\phi}$ by
\begin{equation}
M\sbs{\phi}=c\sbs{s}\sps{2}\left(1/\omega\sbs{\kappa} - 1/2\right)\Delta t,
\end{equation}
and typically we let all other relaxation rates be equal to unity. Finally, what follows are the algorithmic steps taken to solve the LBE across one time step. This process is then marched in time for dynamic tracking of interfaces.
\begin{itemize}
\item Compute pre-collision raw moments from distribution functions through the transformation
\begin{equation}
\bm{\kappa'} = \tensr{P}\bm{f} \nonumber
\end{equation}

\item Compute pre-collision central moments from raw moments using
\begin{equation}
\bm{\kappa} = \tensr{F}\bm{\kappa'} \nonumber
\end{equation}

\item Perform collision by relaxing the central moments towards equilibrium central moments
\begin{equation}
\bm{\tilde{\kappa}} = \bm{\kappa} -  \bm{\Lambda\sbs{\kappa}} \left( \bm{\kappa} - \bm{\kappa\sps{eq}}\right) \nonumber
\end{equation}

\item Compute post-collision raw moments from central moments via
\begin{equation}
\bm{\tilde{\kappa}'} = \tensr{F}\sps{-1}\bm{\tilde{\kappa}} \nonumber
\end{equation}

\item Compute post-collision distribution functions from raw moments through the mapping
\begin{equation}
\bm{\tilde{f}} = \tensr{P}\sps{-1}\bm{\tilde{\kappa}'} \nonumber
\end{equation}

\item Perform streaming via
\begin{equation}
f\sbs{\alpha}(\bm{x},t+\Delta t) = \tilde{f}\sbs{\alpha}(\bm{x}-\bm{e}\sbs{\alpha}\Delta t, t)  \nonumber
\end{equation}
where $\alpha = 0,1,2\dots8$.

\item Update the order parameter $\phi$ by computing the zeroth moment of $f\sbs{\alpha}$ as
\begin{equation}
\phi= \sum\sbs{\alpha=0}\sps{8}f\sbs{\alpha}.
\end{equation}

\end{itemize}

\section{Central moment LBM for Two-Fluid Hydrodynamics Based on Navier-Stokes Equations with Variable Surface Tension}\label{app:NSE}
For handling general cases in multiphase flows with high density ratios in LBM, a pressure-based formulation achieved via a transformation of the distribution functions~\cite{he1999lattice}, which is further extended for a central moment-based collision model~\cite{hajabdollahi2021central} is adopted to simulate two-fluid hydrodynamics with variable surface tension effects. Then, the discrete distribution functions $g\sbs{\alpha}$, where $\alpha = 0, 1, \ldots, 8$ are solved to compute the velocity field $\bm{u}$ and pressure $P$ (rather than the density $\rho$, which evolves according to the order parameter $\phi$ in an affine way as noted as the end of this section). In order to develop a central moment LBM for $g\sbs{\alpha}$, we first define the discrete raw moments and central moments, respectively, of $g\sbs{\alpha}$, its equilibria $g\sbs{\alpha}\sps{eq}$, and the source terms $S\sbs{\alpha}$, which account for surface tension and other forces, of order ($m+n$) as
\begin{subequations}
\begin{equation}
\qquad \left( \begin{array}{c}\eta'\sbs{mn}\\[2mm]   \eta'^{\;\sss{eq}}\sbs{mn}\\[2mm] \sigma'\sbs{mn}\end{array} \right)  = \sum\sbs{\alpha = 0}\sps{8} \left( \begin{array}{c}g\sbs{\alpha} \\[2mm]   g\sbs{\alpha}\sps{eq} \\[2mm] S\sbs{\alpha} \end{array} \right)  e\sbs{\alpha x}\sps{m}   e\sbs{\alpha y}\sps{n},\nonumber
\label{rm}
\end{equation}
and
\begin{equation}
\qquad \left( \begin{array}{c}\eta\sbs{mn} \\[2mm]   \eta\sbs{mn}\sps{eq} \\[2mm] \sigma\sbs{mn}\end{array} \right)  = \sum\sbs{\alpha = 0}\sps{8} \left( \begin{array}{c}g\sbs{\alpha} \\[2mm]  g\sbs{\alpha}\sps{eq} \\[2mm] S\sbs{\alpha} \end{array} \right) (e\sbs{\alpha x}-u\sbs{x})\sps{m}  ( e\sbs{\alpha y}-u\sbs{y})\sps{n}.\nonumber
\label{cm}
\end{equation}
\end{subequations}
Then, we list the nine independent discrete raw moments and central moments supported by the D2Q9 lattice as the vectors $\bm{\eta'}$ and $\bm{\eta}$ as
\begin{subequations}
\begin{eqnarray}
\qquad \bm{\eta'} \! \! \! &=& \! \! \! ( \eta'\sbs{00}, \eta'\sbs{10},\eta'\sbs{01}, \eta'\sbs{20}, \eta'\sbs{02}, \eta'\sbs{11},\eta'\sbs{21}, \eta'\sbs{12},\eta'\sbs{22} )\sps{\top},\label{eqn:4a}\nonumber \\[3mm]
\qquad \bm{\eta} \! \! \! &=& \! \! \! ( \eta\sbs{00},\eta\sbs{10}, \eta\sbs{01}, \eta\sbs{20}, \eta\sbs{02}, \eta\sbs{11}, \eta\sbs{21}, \eta\sbs{12}, \kappa\sbs{22} )\sps{\top}.\nonumber
\end{eqnarray}
\end{subequations}
Similarly, we can group all the central moments of the equilibria and the source terms as the following vectors
\begin{subequations}
\begin{eqnarray}
\qquad \bm{\eta}\sps{eq} \! \! \! &=& \! \! \! ( \eta\sbs{00}\sps{eq},\eta\sbs{10}\sps{eq}, \eta\sbs{01}\sps{eq}, \eta\sbs{20}\sps{eq}, \eta\sbs{02}\sps{eq}, \eta\sbs{11}\sps{eq}, \eta\sbs{21}\sps{eq}, \eta\sbs{12}\sps{eq}, \kappa\sbs{22}\sps{eq} )\sps{\top}.\nonumber\\[3mm]
\qquad \bm{\sigma} \! \! \! &=& \! \! \! ( \sigma\sbs{00},\sigma\sbs{10}, \sigma\sbs{01}, \sigma\sbs{20}, \sigma\sbs{02}, \sigma\sbs{11}, \sigma\sbs{21}, \sigma\sbs{12}, \sigma\sbs{22} )\sps{\top}.\nonumber
\end{eqnarray}
\end{subequations}

Next, to construct the collision step based on central moments, we match the discrete central moments of equilibria with the continuous central moments of the modified continuous Maxwell distribution that accounts for the pressure-based transformation indicated above. Then, we get~\cite{hajabdollahi2021central}
\begin{gather}
\qquad \eta\sbs{00}\sps{eq} = P, \qquad
\eta\sbs{10}\sps{eq} = \psi(\rho) u\sbs{x},\qquad
\eta\sbs{01}\sps{eq} = \psi(\rho) u\sbs{y},\nonumber \\[2mm]
\qquad \eta\sbs{20}\sps{eq} = c\sbs{s}\sps{2}P + \psi(\rho)u\sbs{x}\sps{2},\qquad
\eta\sbs{02}\sps{eq} = c\sbs{s}\sps{2}P + \psi(\rho)u\sbs{y}\sps{2},\qquad
\eta\sbs{11}\sps{eq} = \psi(\rho)\ux\uy,\nonumber  \\[2mm]
\qquad \eta\sbs{21}\sps{eq} = -\psi(\rho)(\ux\sps{2} + c\sbs{s}\sps{2})\uy,\qquad
\eta\sbs{12}\sps{eq} = -\psi(\rho)(\uy\sps{2} + c\sbs{s}\sps{2})\ux, \nonumber  \\[2mm]
\eta\sbs{22}\sps{eq} = c\sbs{s}\sps{6}\rho + \psi(\rho)(\ux\sps{2}+c\sbs{s}\sps{2})(\uy\sps{2}+c\sbs{s}\sps{2}) ,\nonumber
\end{gather}
where $\psi(\rho)=P-\rho c_s^2$.
Similarly, the discrete central moments of source terms obtained by an analogous matching principle read as~\cite{hajabdollahi2021central}
\begin{gather}
\qquad \sigma\sbs{00} = \Gamma\sps{p}\sbs{00}, \qquad
\sigma\sbs{10} = c\sbs{s}\sps{2} F\sbs{tx} - u\sbs{x} \Gamma\sps{p}\sbs{00},\qquad
\sigma\sbs{01} = c\sbs{s}\sps{2} F\sbs{ty} - u\sbs{y} \Gamma\sps{p}\sbs{00},\nonumber \\[2mm]
\qquad \sigma\sbs{20} = 2  c\sbs{s}\sps{2}   F\sbs{px}  u\sbs{x} +   (\ux\sps{2} + c\sbs{s}\sps{2})  \Gamma\sps{p}\sbs{00}     ,\qquad
\sigma\sbs{02} =  2  c\sbs{s}\sps{2}   F\sbs{py}  u\sbs{y} +   (\uy\sps{2} + c\sbs{s}\sps{2})  \Gamma\sps{p}\sbs{00}  ,\nonumber \\[2mm]
\sigma\sbs{11} = \cscs (F\sbs{px}\uy + F\sbs{py} \ux) + \ux\uy\Gamma\sbs{00}\sps{p} , \nonumber  \\[2mm]
\qquad \sigma\sbs{21} = 0,\qquad
\sigma\sbs{12} =  0 ,\qquad
\sigma\sbs{22} = 0 ,\nonumber
\end{gather}
where $F\sbs{px}=-\partial_x\psi(\rho)$, $F\sbs{py}=-\partial_y\psi(\rho)$, and $\Gamma\sps{p}\sbs{00} = F\sbs{px} u\sbs{x} + F\sbs{py} u\sbs{y}$. In addition, the total force components are $F\sbs{tx}=F\sbs{sx}+F\sbs{ext,x}$ and $F\sbs{ty}=F\sbs{sy}+F\sbs{ext,y}$, where $\bm{F}_s=(F\sbs{sx}, F\sbs{sy})$ and $\bm{F}_{ext}=(F\sbs{ext,x}, F\sbs{ext,y})$ are the local surface tension force (see Eq.~(\ref{Fs}) which accounts for its capillary and Marangoni components that vary locally based on the interfacial surfactant concentration field) and any external force, respectively.

In order to construct the collision step based on central moments for hydrodynamics, for generality and improved stability, we need to segregate the emergent shear viscosity effects from that of bulk viscosity, where the latter is related to the relaxation of the trace of the second order moments under collision. In other words, the collision model should relax $\eta\sbs{20}+\eta\sbs{02}$ and $\eta\sbs{20}-\eta\sbs{02}$ at different rates, where the former is related to the bulk viscosity and the latter to the shear viscosity, rather than relaxing $\eta\sbs{20}$ and $\eta\sbs{02}$ separately. To accomplish this, we introduce the combined second order moment variables
\begin{eqnarray}
\eta\sbs{2s} &=& \eta\sbs{20}+\eta\sbs{02}, \qquad \eta\sbs{2s}\sps{eq} = \eta\sbs{20}\sps{eq}+\eta\sbs{02}\sps{eq},\qquad \sigma\sbs{2s} = \sigma\sbs{20}+\sigma\sbs{02}, \nonumber\\
\eta\sbs{2d} &=& \eta\sbs{20}-\eta\sbs{02}, \qquad \eta\sbs{2d}\sps{eq} = \eta\sbs{20}\sps{eq}-\eta\sbs{02}\sps{eq},\qquad \sigma\sbs{2d} = \sigma\sbs{20}-\sigma\sbs{02},\label{eq:combinedforms}
\end{eqnarray}
and define the following vectors each containing nine independent central moments where the second order diagonal components are replaced with the above combinations:
\begin{subequations}
\begin{eqnarray}
\qquad \bm{\hat{\eta}} \! \! \! &=& \! \! \! ( \eta\sbs{00},\eta\sbs{10}, \eta\sbs{01}, \eta\sbs{2s}, \eta\sbs{2d}, \eta\sbs{11}, \eta\sbs{21}, \eta\sbs{12}, \kappa\sbs{22} )\sps{\top}.\nonumber\\[3mm]
\qquad \bm{\hat{\eta}}\sps{eq} \! \! \! &=& \! \! \! ( \eta\sbs{00}\sps{eq},\eta\sbs{10}\sps{eq}, \eta\sbs{01}\sps{eq}, \eta\sbs{2s}\sps{eq}, \eta\sbs{2d}\sps{eq}, \eta\sbs{11}\sps{eq}, \eta\sbs{21}\sps{eq}, \eta\sbs{12}\sps{eq}, \kappa\sbs{22}\sps{eq} )\sps{\top}.\nonumber\\[3mm]
\qquad \bm{\hat{\sigma}} \! \! \! &=& \! \! \! ( \sigma\sbs{00},\sigma\sbs{10}, \sigma\sbs{01}, \sigma\sbs{2s}, \sigma\sbs{2d}, \sigma\sbs{11}, \sigma\sbs{21}, \sigma\sbs{12}, \sigma\sbs{22} )\sps{\top}.\nonumber
\end{eqnarray}
\end{subequations}
Then, the vectors that were defined earlier (which involved bare forms of diagonal second order moments) with these vectors (which involve combined forms of second order moments and identified with the `hat' over their notations) can be related formally as
\begin{equation}
\bm{\hat{\eta}} = \tensr{B}\bm{\eta}, \qquad \bm{\hat{\eta}}\sps{eq} = \tensr{B}\bm{\eta}\sps{eq}, \qquad \bm{\hat{\sigma}} = \tensr{B}\bm{\sigma},\label{eq:blocktransformation}
\end{equation}
where $\tensr{B}$ is a block diagonal matrix that transforms such bare moments into combined moments. Based on these considerations, we can then write the central moment LBM for two-fluid hydrodynamics with variable surface tension effects as
\begin{equation}
g\sbs{\alpha}(\bm{x}+\bm{e}\sbs{\alpha}\Delta t, t+\Delta t) - g\sbs{\alpha}(\bm{x}, t) = \tensr{P}\sps{\um 1} \tensr{F}\sps{\um 1} \tensr{B}\sps{\um 1} \left\{\bm{\Lambda\sbs{\eta}}\left[ \bm{\hat{\eta}}\sps{eq} - \bm{\hat{\eta}}\right] + \left(1-\frac{\bm{\Lambda\sbs{\eta}}}{2}\right) \bm{\hat{\sigma}}\Delta t \right\},\label{eq:centralmomentLBE2fluidflow}
\end{equation}
where $\bm{\Lambda\sbs{\eta}}$ is the relaxation rate matrix given by $\bm{\Lambda\sbs{\eta}} = \mbox{diag}(1, 1, 1, \omega\sbs{\zeta}, \omega\sbs{\nu}, \omega\sbs{\nu},1 , 1, 1)$. Note that the shear viscosity $\nu$ and bulk viscosity $\zeta$ are related to the respective relaxation rates via
\begin{equation}
\nu=c\sbs{s}\sps{2}\left(1/\omega\sbs{\nu} - 1/2\right)\Delta t, \qquad \zeta=c\sbs{s}\sps{2}\left(1/\omega\sbs{\zeta} - 1/2\right)\Delta t.
\end{equation}
Here, typically we let all other relaxation rates be equal to unity, based on simplicity and stability considerations. Note that while Eq.~(\ref{eq:blocktransformation}) involves a formal transformation involving multiplication of vectors with a block matrix, in practical implementations, no such multiplications are necessary as these mappings simply imply that we replace component-wise the two second order diagonal bare moments with the combined forms as shown in Eq.~(\ref{eq:combinedforms}) before collision; then following collision, the post-collision components (referred with `tilde' over symbols in what follows) in their bare forms can be retrieved from their combined forms via $\tilde{\eta}\sbs{20}=(\tilde{\eta}\sbs{2s}+\tilde{\eta}\sbs{2d})/2$ and $\tilde{\eta}\sbs{02}=(\tilde{\eta}\sbs{2s}-\tilde{\eta}\sbs{2d})/2$, which amount to applying the inverse operator $\tensr{B}^{-1}$ in Eq.~(\ref{eq:centralmomentLBE2fluidflow}).

The algorithmic steps for implementing Eq.~(\ref{eq:centralmomentLBE2fluidflow}) are outlined in what follows, which are marched in time for computing the flow fields, viz., fluid velocity and pressure.
\begin{itemize}
\item Compute pre-collision raw moments from distribution functions via
\begin{equation}
\bm{\eta'} = \tensr{P}\bm{g} \nonumber
\end{equation}

\item Compute pre-collision central moments from raw moments through
\begin{equation}
\bm{\eta} = \tensr{F}\bm{\eta'} \nonumber
\end{equation}

\item Perform collision by relaxing the central moments towards equilibrium central moments
First, combine second order diagonal central moments using
\begin{equation*}
\bm{\hat{\eta}} = \tensr{B}\bm{\eta}, \qquad \bm{\hat{\eta}}\sps{eq} = \tensr{B}\bm{\eta}\sps{eq}, \qquad \bm{\hat{\sigma}} = \tensr{B}\bm{\sigma},
\end{equation*}
and then relax them under collision with updates from the source terms as
\begin{equation}
\bm{\tilde{\hat{\eta}}} = \bm{\eta} + \bm{\Lambda\sbs{\eta}} \left( \bm{\hat{\eta}}\sps{eq} - \bm{\hat{\eta}} \right) + \left(1-\frac{\bm{\Lambda\sbs{\eta}}}{2}\right)\bm{\hat{\sigma}}\Delta t,  \nonumber
\end{equation}
which is followed by retrieving the post-collision central moments in their bare forms using
\begin{equation*}
\bm{\tilde{\eta}} = \tensr{B}^{-1}\bm{\tilde{\hat{\eta}}}
\end{equation*}

\item Compute post-collision raw moments from central moments through
\begin{equation}
\bm{\tilde{\eta}'} = \tensr{F}\sps{-1}\bm{\tilde{\eta}} \nonumber
\end{equation}

\item Compute post-collision distribution functions from raw moments via
\begin{equation}
\bm{\tilde{g}} = \tensr{P}\sps{-1}\bm{\tilde{\eta}'} \nonumber
\end{equation}

\item Perform streaming via
\begin{equation}
g\sbs{\alpha}(\bm{x},t+\Delta t) = \tilde{g}\sbs{\alpha}(\bm{x}-\bm{e}\sbs{\alpha}\Delta t, t),  \nonumber
\end{equation}
where $\alpha = 0,1,2\dots8$.

\item Update the pressure and fluid velocity by computing the zeroth moment and first order moments, respectively, of $g\sbs{\alpha}$ as
\begin{equation}
P = \sum\sbs{\alpha=0}\sps{8}g\sbs{\alpha} + \frac{1}{2}\sigma\sbs{00}\Delta t, \qquad c\sbs{s}\sps{2} \rho \bm{u} = \sum\sbs{\alpha=0}\sps{8} \bm{e}\sbs{\alpha} g\sbs{\alpha} + \frac{1}{2}c\sbs{s}\sps{2}\bm{F\sbs{t}}\Delta t,
\end{equation}
and the local density $\rho$ and viscosity $\nu$ are obtained from the order parameter $\phi$ as follows so that they vary smoothly across the interfaces:
\begin{equation}
\rho = \rho\sbs{b}+\left(\frac{\phi-\phi\sbs{b}}{\phi\sbs{a}-\phi\sbs{b}}\right)(\rho\sbs{a}-\rho\sbs{b}), \qquad
\nu = \nu\sbs{b}+\left(\frac{\phi-\phi\sbs{b}}{\phi\sbs{a}-\phi\sbs{b}}\right)(\nu\sbs{a}-\nu\sbs{b})
\end{equation}

\end{itemize}

%\newpage
%\bibliography{bib.bib}{}
%\bibliographystyle{unsrt}

\end{document}